\documentclass[11pt,a4paper]{article}

\usepackage{amsmath}
\usepackage{amssymb}
\usepackage{graphicx}
\usepackage{bm}

\usepackage{mathrsfs}
\usepackage{cite}

\newcommand*\xbar[1]{%
  \hbox{%
    \vbox{%
      \hrule height 0.5pt 
      \kern0.5ex
      \hbox{%
        \kern-0.1em
        \ensuremath{#1}%
        \kern-0.1em
      }%
    }%
  }%
}

\numberwithin{equation}{section}

\begin{document}

\title{The structure of divergences in the higher-derivative\\ supersymmetric $6D$  gauge theory}

\author{
I.L.Buchbinder\footnote{buchbinder@theor.jinr.ru}\ ${}^{a,b,c}$, A.S.Budekhina\footnote{budekhina@theor.jinr.ru}\ ${}^{a}$,
E.A.Ivanov\footnote{eivanov@theor.jinr.ru}\ ${}^{a,d}$, and K.V.Stepanyantz\footnote{stepan@m9com.ru}\ ${}^{e}$\\
\\
${}^a${\small{\em Joint Institute for Nuclear Research}}, {\small{\em  Bogoliubov Laboratory of Theoretical Physics}}\\
{\small{\em 141980, Dubna, Moscow region, Russia}}\\
${}^b${\small{\em Center of Theoretical Physics, Tomsk State
Pedagogical University, 634061, Tomsk, Russia}}\\
${}^c${\small{\em National Research Tomsk State University, 634050,
Tomsk, Russia}}\\
${}^d${\small{\em Moscow Institute of Physics and Technology, 141700 Dolgoprudny, Moscow region, Russia}}\\
${}^e${\small{\em Moscow State University}}, {\small{\em  Faculty of Physics, Department  of Theoretical Physics}}\\
{\small{\em 119991, Moscow, Russia}}\\
}

\maketitle

\begin{abstract}
Using the harmonic superspace approach, we perform a comprehensive
study of the structure of divergences in the higher-derivative $6D$,
${\cal N}=(1,0)$ supersymmetric Yang--Mills theory coupled to the
hypermultiplet in the adjoint representation. The effective action
is constructed in the framework of the superfield background field
method with the help of ${\cal N}=(1,0)$ supersymmetric
higher-derivative regularization scheme which preserves all
symmetries of the theory. The one-loop divergences are calculated in
a manifestly gauge invariant and $6D$, ${\cal N}=(1,0)$
supersymmetric form hopefully admitting a generalization to higher
loops. The $\beta$-function in the one-loop approximation is found
and analyzed. In particular, it is shown that the one-loop
$\beta$-function for an arbitrary regulator function is specified by
integrals of double total derivatives in momentum space, like it happens in $4D,\,
{\cal N}=1$ superfield gauge theories. This points to the potential
possibility to derive the {\it all-loop} NSVZ-like exact
$\beta$-function in the considered theory.
\end{abstract}

\section{Introduction}
\hspace*{\parindent}

The higher derivative field theories attract an increasing interest. In spite of the well known problems with ghosts, they possess a number of  remarkable properties in the quantum domain,
which suggests treating them as useful effective models. There exists a huge literature on the quantum structure of higher derivative theories, mainly in the context
of quantum gravity (see, e.g., a review \cite{Smilga:2017arl} and references therein, as well as recent papers \cite{Buccio:2024hys,Lacombe:2024jac}).

It is well known that the supersymmetric theories in four dimensions have a better ultraviolet behavior compared to their non-supersymmetric counterparts due to
the non-renormalization theorems (see, e.g., \cite{Gates:1983nr,West:1990tg,Buchbinder:1998qv}). This entails remarkable consequences for the structure
of supersymmetric quantum field theories. For instance,  because of absence of divergent quantum corrections to the
superpotential \cite{Grisaru:1979wc}, it becomes possible to relate the renormalization of masses and Yukawa couplings to the renormalization of the chiral matter superfields.
As a result, the mass
anomalous dimension and the Yukawa $\beta$-function can be expressed in terms of the anomalous dimension of the matter superfields. It turns out that the gauge $\beta$-function
can also be related to the
anomalous dimension of the chiral matter superfields by the exact NSVZ equation \cite{Novikov:1983uc,Jones:1983ip,Novikov:1985rd,Shifman:1986zi}.
In the particular case of pure $4D$, ${\cal N}=1$ supersymmetric
Yang--Mills (SYM) theory the NSVZ approach yields the all-order exact expression for the $\beta$-function in the form of a geometric series.
It is natural to expect that supersymmetry in higher-derivative theories is capable to further soften the ultraviolet behavior as compared
to the conventional bosonic higher-derivative theories.

Supersymmetric theories in higher dimensions are under an intensive study at present
(see e.g. \cite{Howe:1983jm,Howe:2002ui,Bossard:2009sy,Bossard:2009mn,Fradkin:1982kf,Marcus:1983bd,Smilga:2016dpe,Bork:2015zaa}),
mainly due to their numerous relations to string/brane stuff. The conventional higher-dimensional Yang--Mills theories
are not renormalizable by power counting because of the dimensionful coupling constant\footnote{Nevertheless, such theories can inherit
some properties of the underlying superstring theory, which may simplify their renormalizaiton structure (see, e.g.,
a  recent paper \cite{Buchbinder:2022rju} and references therein).}. However, it is possible to construct UV
complete supersymmetric theories by introducing higher derivatives in the Lagrangian. Depending on the degree of derivatives,
this  trick allows one to gain the higher-derivative renormalizable or even super-renormalizable theories. For instance, $6D,$ ${\cal N}=(1,0)$ SYM theory
with higher derivatives constructed in \cite{Ivanov:2005qf} is renormalizable by power counting. The one-loop divergences for this theory have been
considered in \cite{Ivanov:2005qf,Casarin:2019aqw}\footnote{See also \cite{Casarin:2017xez,Casarin:2024qdn}.}
by making use of a non-manifestly supersymmetric component formulation and the regularization by dimensional reduction \cite{Siegel:1979wq}.
The manifestly ${\cal N}=(1,0)$ supersymmetric form of the action can be achieved in the framework of harmonic superspace. Originally,
the latter has been developed for the $4D$ case, see
\cite{Galperin:1984av,Galperin:1985bj,Galperin:1985va,Galperin:2001uw,Buchbinder:2001wy,Buchbinder:2016wng}. A generalization
to the case of $6D$ supersymmetric theories was given in
\cite{Howe:1983fr,Howe:1985ar,Zupnik:1986da,Ivanov:2005kz,Buchbinder:2014sna}. The relevant techniques allow one to explicitly
calculate supersymmetric quantum corrections, provided that the regularization preserves supersymmetry at each order of the loop expansion.
For the higher-derivative ${\cal N}=(1,0)$ theory formulated in the harmonic superspace the one-loop divergences were found
in \cite{Buchbinder:2020tnc,Buchbinder:2020ovf}, applying the regularization by dimensional reduction\footnote{For the conventional
$6D$ ${\cal N}=(1,0)$ and ${\cal N}=(1,1)$ supersymmetric theories various calculations of quantum corrections were performed in
\cite{Buchbinder:2016gmc,Buchbinder:2017gbs,Buchbinder:2017xjb,Buchbinder:2018lbd,Buchbinder:2018bhs,Buchbinder:2019gfb,Buchbinder:2019vvc,Buchbinder:2021unt,Buchbinder:2022rju}.}.
Although this regularization is efficient enough and is very often used in the concrete calculations (see, e.g., \cite{Jack:1997sr,Gnendiger:2017pys}
and references therein), it can violate supersymmetry in higher loops. It seems useful to reconsider the previously derived results on the structure
of the effective action in the framework of a regularization which would guarantee unbroken supersymmetry. At present, the only regularization of supersymmetric
gauge theories which meets such a demand is the one through higher covariant derivatives (see, e.g., \cite{Stepanyantz:2019lyo}
and references therein) originally proposed by Slavnov \cite{Slavnov:1971aw,Slavnov:1972sq,Slavnov:1977zf}. For $4D$ supersymmetric theories this regularization
can be formulated both in the ${\cal N}=1$ superfield language \cite{Krivoshchekov:1978xg,West:1985jx} and  using the harmonic superspace \cite{Buchbinder:2015eva}.

When applied to $4D$ supersymmetric gauge theories, the higher covariant derivative regularization makes
it possible to express the $\beta$-function through integrals of double total derivatives in the momentum space
\cite{Smilga:2004zr,Stepanyantz:2011jy,Stepanyantz:2019ihw}. This property leads to the appearance
of the NSVZ expression for the renormalization group functions (RGFs) defined in terms
of the bare couplings or, equally, for the standard RGFs in the HD+MSL
scheme \cite{Kataev:2013eta,Stepanyantz:2020uke}. It can be derived by summing the singular contributions
(which appear when the double total derivatives act on the inverse-squared momenta)
\cite{Stepanyantz:2019lfm,Stepanyantz:2020uke} and taking into account
the non-renormalization of the triple gauge ghost vertices \cite{Stepanyantz:2016gtk}.
All these general statements have been confirmed by various multiloop calculations,
see, e.g., \cite{Shakhmanov:2017soc,Kazantsev:2018nbl,Kuzmichev:2019ywn,Aleshin:2020gec,Kuzmichev:2021yjo,Kuzmichev:2021lqa,Aleshin:2022zln,Shirokov:2023jya}.
In particular, the factorization into integrals of double derivatives can be observed already in the one-loop approximation \cite{Aleshin:2016yvj}.
Note that for an arbitrary form of the higher derivative regulator this result is non-trivial and actually reflects s
some deep features of quantum $4D$ supersymmetric theories. This is the basic reason  why we believe that the possibility
of similar factorization for higher-dimensional supersymmetric theories could shed more light on their
underlying quantum structure.

In this paper we consider $6D$, ${\cal N}=(1,0)$ supersymmetric theory with higher derivatives
proposed in \cite{Ivanov:2005qf}, with the action modified by adding the matter hypermultiplet in the adjoint representation of gauge group.
The $6D$, ${\cal N}=(1,0)$ supersymmetric theories are in general anomalous \cite{Kuzenko:2017xgh}.
However, in the presence of hypermultiplet the theory under consideration is anomaly-free \cite{Smilga:2006ax}.
We will regularize it by higher covariant derivatives
with the help of the manifestly ${\cal N}=(1,0)$ supersymmetric technique based on the $6D$ version of harmonic
superspace \cite{Howe:1983fr,Howe:1985ar,Zupnik:1986da,Ivanov:2005kz,Buchbinder:2014sna}.
Using the regularization constructed we calculate the one-loop divergences and demonstrate that the $\beta$-function is determined by integrals of the double total derivatives for an arbitrary choice of the regulator function. The resemblance of the resulting expression to the analogous expression in $4D$ case suggests that some other features of $4D$ supersymmetric theories could also be valid in $6D$ case. In particular, we guess that the $\beta$-function of the model under consideration is given by integrals of the double total derivatives to any loop and so can be brought in the NSVZ-like form.

The paper is organized as follows. In Sect. \ref{Section_Model} the basic features of the theory under consideration
are overviewed. In particular, in Subsect. \ref{Subsection_Harmonic_Superspace} we briefly recall some information about the harmonic superspace.
This technique is used in Subsect. \ref{Subsection_Higher_Derivative_Theory} for constructing the manifestly ${\cal N}=(1,0)$ supersymmetric
and gauge invariant action. The theory is regularized by higher covariant derivatives in Subsect. \ref{Subsection_HD_Regularization}.
The quantization procedure is described in Subsect. \ref{Subsection_Quantization}. Sect. \ref{Section_One_loop} is devoted to the calculation
of the one-loop divergent quantum corrections. In particular, we demonstrate that the one-loop $\beta$-function can be written
as an integral of the double total derivatives in the momentum space. Possible generalizations of this result and its consequences
for the higher-order quantum corrections are discussed in Sect. \ref{Section_Higher_Orders}. The results of the paper
are briefly summarized in Conclusion, while some technical details of the calculations are expounded in Appendix.

\section{Higher-derivative $6D$, ${\cal N}=(1,0)$ supersymmetric Yang--Mills theory}
\label{Section_Model}

\subsection{Harmonic superspace}
\label{Subsection_Harmonic_Superspace}
\hspace*{\parindent}

The manifestly ${\cal N}=(1,0)$ supersymmetric description of $6D$ theories is achieved
in harmonic superspace (see, e.g., \cite{Bossard:2015dva}). It is parametrized by the ordinary
space-time coordinates $x^M$ (with $M=0,\ldots,5$), the anticommuting left-handed
spinors $\theta^a_i$ (with $a=1,\ldots, 4$ and $i=1,2$), and the harmonic variables $u^{\pm i}$, such that

\begin{equation}
u^-_i = (u^{+i})^*,\qquad u^{+i} u^-_i = 1,\qquad  u_i^\pm = \varepsilon_{ij} u^{\pm j}.
\end{equation}

\noindent
The harmonic superspace contains a subspace which is closed under supersymmetry transformations and involves half the original Grassmann coordinates.
This (analytic) subspace is parametrized by the coordinate set

\begin{equation}
x_A^M = x^M + \frac{i}{2} \theta^- \gamma^M \theta^+, \qquad
\theta^{+a} = u^{+}_i \theta^{a i}, \qquad u^\pm_i,
\end{equation}

\noindent
where the $6D$ $\gamma$-matrices are denoted by $\gamma^M$.

Using the harmonic variables one can construct the harmonic derivatives

\begin{equation}
D^{++} = u^{+i} \frac{\partial}{\partial u^{-i}},\qquad D^{--}
= u^{-i} \frac{\partial}{\partial u^{+i}},\qquad D^0 = u^{+i} \frac{\partial}{\partial u^{+i}} - u^{-i} \frac{\partial}{\partial u^{-i}},
\end{equation}

\noindent
and the harmonic spinor covariant derivatives

\begin{equation}
D_a^+ = u^+_i D^i_a,\qquad D_a^- = u^-_i D^i_a.
\end{equation}

\noindent
By definition, an analytic superfield $\phi$ satisfies the Grassmann analyticity constraint $D_a^+ \phi = 0$.

For constructing the $6D$, ${\cal N}=(1,0)$ supersymmetric actions we use the superspace integration measures.
In the notation adopted in this paper, they are defined as

\begin{equation}
\int d^{14}z \equiv \int d^6x\,d^8\theta,\qquad \int d\zeta^{(-4)} \equiv \int d^6x\,d^4\theta^+\,du,
\end{equation}

\noindent
where $z\equiv (x^M,\theta^a_i)$. The integration measures are related as

\begin{equation}
\int d\zeta^{(-4)} (D^+)^4 = \int d^{14}z\, du,
\end{equation}

\noindent
where

\begin{equation}
(D^+)^4 \equiv - \frac{1}{24} \varepsilon^{abcd} D_a^+ D_b^+ D_c^+ D_d^+.
\end{equation}

\subsection{The higher-derivative theory}
\label{Subsection_Higher_Derivative_Theory}
\hspace*{\parindent}

We consider $6D$, ${\cal N}=(1,0)$ higher-derivative SYM theory proposed in \cite{Ivanov:2005qf}, to which we add the hypermultiplet in the adjoint representation.
This theory is described by the action

\begin{equation}\label{Action_Of_The_Theory}
S = \pm \frac{1}{2e_0^2} \mbox{tr} \int d\zeta^{(-4)} (F^{++})^2 - \frac{2}{e_0^2}\,\mbox{tr} \int d\zeta^{(-4)} \widetilde{q^+} \nabla^{++} q^+.
\end{equation}

\noindent
Note that the sign of the usual Yang-Mills action is chosen so that the energy is positive.
However, this requirement cannot in general be satisfied for theories with higher derivatives.
That is why, following \cite{Buchbinder:2020tnc,Buchbinder:2020ovf}, in the gauge part of the action (\ref{Action_Of_The_Theory})
we admit both signs. Below the upper and lower signs will always correspond to the plus and minus signs
in eq. (\ref{Action_Of_The_Theory}), respectively.

According to reasonings of \cite{Smilga:2006ax}, the theory (\ref{Action_Of_The_Theory}) is anomaly-free
due to the presence of the hypermultiplet in the adjoint representation.
Moreover, we will see that the theory with hypermultiplet avoids one-loop quadratic divergences,
which is very important for constructing the higher covariant derivative regularization
in $6D$ space-time\footnote{The hypermultiplet term in (\ref{Action_Of_The_Theory}) breaks ${\cal N}=2$ superconformal
symmetry respected by the first term. As argued in \cite{Ivanov:2005kz}, there exist no superconformally
invariant $6D$ hypermultiplet actions. On the other hand, nothing forbids adding
some higher-derivative hypermultiplet terms to (\ref{Action_Of_The_Theory})
(e.g., like those discussed in \cite{Ivanov:2005kz}). We limit our consideration to (\ref{Action_Of_The_Theory}) for the minimality reasons.}.

In eq. (\ref{Action_Of_The_Theory}) $e_0$ is the dimensionless bare coupling constant (hereafter,
all bare couplings are marked by the subscript $0$). The gauge superfield $V^{++} = e_0 V^{++A} t^A$
and the hypermultiplet superfield $q^+ = e_0 q^{+A} t^A$ satisfy the analyticity conditions

\begin{equation}
D_a^+ V^{++} = 0,\qquad D_a^+ q^+ = 0.
\end{equation}

\noindent
The generators of the gauge group $t^A$ are taken in the fundamental representation for which the normalization condition and the commutation relations can be written as

\begin{equation}
\mbox{tr}(t^A t^B) = \frac{1}{2} \delta^{AB},\qquad [t^A,t^B] = i f^{ABC} t^C,
\end{equation}

\noindent
where $f^{ABC}$ are real and totally antisymmetric structure constants.

The generalized conjugation is denoted by tilde, the gauge superfield being real
with respect to it, $\widetilde{V^{++}} = V^{++}$. The conjugated hypermultiplet superfield $\widetilde{q^+}$ is also analytic.

In the harmonic approach the covariant gauge field strength is the analytic superfield $F^{++}$ defined as

\begin{equation}
F^{++} \equiv (D^+)^4 V^{--},
\end{equation}

\noindent
where the (non-analytic) superfield

\begin{equation}\label{V--}
V^{--}(z,u) \equiv \sum\limits_{n=1}^\infty (-i)^{n+1} \int du_1\ldots du_n \frac{V^{++}(z,u_1) V^{++}(z,u_2)\ldots V^{++}(z,u_n)}{(u^+ u_1^+) (u_1^+ u_2^+)\ldots (u_n^+ u^+)}
\end{equation}

\noindent
satisfies the zero curvature constraint

\begin{equation}
D^{++} V^{--} - D^{--} V^{++} + i[V^{++},V^{--}] = 0.
\end{equation}

\noindent
The harmonic covariant derivatives of hypermultiplet are written as

\begin{equation}\label{Covariant_Detivatives}
\nabla^{\pm\pm} q^+  = D^{\pm\pm} q^+ + i [V^{\pm\pm},q^+].
\end{equation}

The theory (\ref{Action_Of_The_Theory}) is invariant under the gauge transformations

\begin{equation}\label{Gauge_Transformations}
V^{++} \to e^{i\lambda} V^{++} e^{-i\lambda} - i e^{i\lambda} D^{++} e^{-i\lambda}, \qquad q^+ \to e^{i\lambda} q^+ e^{-i\lambda},
\end{equation}

\noindent
where the parameter $\lambda = e_0 \lambda^A t^A$ is an analytic superfield, such that $\widetilde{\lambda^A} = \lambda^A$.
Under these transformations other quantities defined above behave as:

\begin{eqnarray}
&& V^{--} \to e^{i\lambda} V^{--} e^{-i\lambda} - i e^{i\lambda} D^{--} e^{-i\lambda},\qquad F^{++} \to e^{i\lambda} F^{++} e^{-i\lambda}, \qquad\nonumber\\
&& \nabla^{\pm\pm} q^+ \to e^{i\lambda} \nabla^{\pm\pm} q^+ e^{-i\lambda},\qquad\qquad\qquad\ \, \widetilde{q^+} \to e^{i\lambda} \widetilde{q^+} e^{-i\lambda}.\label{OtherGau}
\end{eqnarray}

\subsection{Regularization by higher-order covariant derivatives}
\label{Subsection_HD_Regularization}
\hspace*{\parindent}

The higher covariant-derivative regularization is introduced by inserting a term with higher derivatives into the action (\ref{Action_Of_The_Theory}).
For constructing such a term it is necessary to take into account the identity which is satisfied for an arbitrary analytic superfield $\phi$

\begin{equation}
\frac{1}{2} (D^+)^4 (D^{--})^2 \phi = \partial^2\phi.
\end{equation}

\noindent
Obviously, the needed higher derivative term should be gauge invariant. So, in order to construct it, we should use the analogous operator containing covariant derivatives,

\begin{equation}\label{Box_Definition}
\Box \equiv \frac{1}{2} (D^+)^4 (\nabla^{--})^2.
\end{equation}

\noindent
Then the sum of the original action and the higher derivative term can be written in the form

\begin{equation}\label{Action_Of_Regularized_Theory}
S_{\mbox{\scriptsize reg}} = \pm \frac{1}{2e_0^2} \mbox{tr} \int d\zeta^{(-4)} F^{++} R\Big(\frac{\Box}{\Lambda^2}\Big) F^{++} - \frac{2}{e_0^2}\,\mbox{tr} \int d\zeta^{(-4)} \widetilde{q^+} \nabla^{++} q^+,
\end{equation}

\noindent
where $\Lambda$ is a dimensionful parameter which plays the role of an ultraviolet cutoff. The regulator function $R(x)$ should increase at $x\to\infty$
and satisfy the condition $R(0)=1$ (this condition is required in order to recover the original theory (\ref{Action_Of_The_Theory}) in the limit $\Lambda\to \infty$).
It is easy to verify that the action (\ref{Action_Of_Regularized_Theory}) is invariant under the gauge transformations (\ref{Gauge_Transformations}), (\ref{OtherGau}).

However, according to \cite{Slavnov:1977zf}, after adding the higher-derivative term to the classical action,
divergences still survive in the one-loop approximation. To regularize these residual divergences, one is led to insert the Pauli--Villars fields with a mass parameter
in the generating functional. The details of this procedure for the theory we deal with will be discussed in the next section.

\subsection{Quantization}
\label{Subsection_Quantization}
\hspace*{\parindent}

For quantizing gauge theories it is convenient to use the background field method \cite{DeWitt:1965jb,DeWitt}
(see also \cite{Kallosh,Arefeva:1974jv,Abbott:1980hw,Abbott:1981ke}), because it allows constructing
the manifestly gauge invariant effective action. The first step of background field method is the background-quantum splitting. In harmonic superspace
such a splitting is carried out as follows \cite{Buchbinder:2001wy,Buchbinder:1997ya}

\begin{equation}\label{Splitting}
V^{++} = \bm{V}^{++} + v^{++},
\end{equation}

\noindent
where $\bm{V}^{++}$ and $v^{++}$ are the background and quantum gauge superfields, respectively. In this case the original gauge invariance (\ref{Gauge_Transformations})
amounts to two different symmetries. The quantum gauge symmetry,

\begin{equation}\label{Quantum_Symmetry}
\bm{V}^{++} \to \bm{V}^{++},\quad v^{++} \to e^{i\lambda} (v^{++} +\bm{V}^{++})e^{-i\lambda} - \bm{V}^{++} - i e^{i\lambda} D^{++} e^{-i\lambda},\quad q^+ \to e^{i\lambda} q^+\,,
\end{equation}

\noindent
is broken by the gauge-fixing procedure down to the BRST invariance. The second, background transformations,

\begin{equation}\label{Background_Symmetry}
\bm{V}^{++} \to e^{i\lambda} \bm{V}^{++} e^{-i\lambda} -i e^{i\lambda}D^{++} e^{-i\lambda},\qquad v^{++} \to e^{i\lambda} v^{++} e^{-i\lambda},\qquad q^+ \to e^{i\lambda} q^+\,,
\end{equation}

\noindent
remain the manifest symmetry of the effective action. Certainly, this occurs only providing we choose the gauge-fixing term which preserves this symmetry.
For instance, it is convenient to fix a gauge by adding the term

\begin{equation}\label{Gauge_Fixing_Term}
S_{\mbox{\scriptsize gf}} = \mp \frac{1}{2e_0^2\xi_0} \mbox{tr} \int d^{14}z\,du_1\,du_2\, e^{i\bm{b}_1} e^{-i\bm{b}_2} (\bm{\nabla}^{++}_2 v^{++}_2) e^{i\bm{b}_2} e^{-i\bm{b}_1} \frac{(u_1^- u_2^-)}{(u_1^+ u_2^+)^3} R\Big(\frac{\bm{\Box}_1}{\Lambda^2}\Big) \bm{\Box}_1 \bm{\nabla}^{++}_1 v^{++}_1,
\end{equation}

\noindent
where $\xi_0$ is the bare gauge fixing parameter, and the subscripts denote different sets of harmonic variables, e.g., $v^{++}_1 \equiv v^{++}(z,u_1)$, etc.
The background covariant derivatives

\begin{equation}
\bm{\nabla}^{\pm\pm} v^{++} \equiv D^{\pm\pm} v^{++} + i [\bm{V}^{\pm\pm}, v^{++}]
\end{equation}

\noindent
are constructed in terms of the background gauge superfield $\bm{V}^{\pm\pm}$ (unlike the derivative (\ref{Covariant_Detivatives})
which includes the total gauge superfield $V^{\pm\pm}$).
The operator $\bm{\Box}$ in eq. (\ref{Gauge_Fixing_Term}) is defined exactly as the operator $\Box$ in eq. (\ref{Box_Definition}),
but the derivatives $\nabla^{--}$ should be replaced
by the background-group covariant derivatives $\bm{\nabla}^{--}$. The superfield $\bm{b}$ present
in the gauge-fixing term (\ref{Gauge_Fixing_Term}) is the so called bridge,
it is related to the background superfields $\bm{V}^{\pm\pm}$ by the equation

\begin{equation}
\bm{V}^{++} = - i e^{i\bm{b}} D^{++} e^{-i\bm{b}},\qquad \bm{V}^{--} = - i e^{i\bm{b}} D^{--} e^{-i\bm{b}}.
\end{equation}

\noindent
The gauge-fixing term (\ref{Gauge_Fixing_Term}) is invariant under the transformations (\ref{Background_Symmetry})
which act on the bridge as $e^{i\bm{b}} \to e^{i\lambda} e^{i\bm{b}} e^{-i\tau}$, where $\tau=\tau(z)$ is independent of the harmonic variables.

The action of Faddeev--Popov ghosts corresponding to the gauge-fixing term (\ref{Gauge_Fixing_Term}) has the standard form

\begin{equation}
S_{\mbox{\scriptsize FP}} = \frac{2}{e_0^2}\,\mbox{tr} \int d\zeta^{(-4)} \bm{\nabla}^{++} b \Big(\bm{\nabla}^{++} c + i[v^{++}, c]\Big),
\end{equation}

\noindent
where $b$ and $c$ are anticommuting analytic superfields in the adjoint representation. When using the background field
method for supersymmetric theories,
it is also necessary to insert the Nielsen--Kallosh determinant. For the case at hand it can be written as

\begin{equation}\label{Nielsen_Kallosh1}
\Delta_{\mbox{\scriptsize NK}} = \mbox{Det}^{1/2}\Big[\bm{\Box}^2 R\Big(\frac{\bm{\Box}}{\Lambda^2}\Big)\Big]\, \int D\varphi\, \exp (iS_{\mbox{\scriptsize NK}}),
\end{equation}

\noindent
where $\varphi$ is a commuting analytic superfield in the adjoint representation,

\begin{equation}\label{Nielsen_Kallosh2}
S_{\mbox{\scriptsize NK}} = - \frac{1}{e_0^2}\, \mbox{tr} \int d\zeta^{(-4)} (\bm{\nabla}^{++}\varphi)^2,
\end{equation}

\noindent
and the function $R(x)$ appears due to its presence in the expression (\ref{Gauge_Fixing_Term}).
The corresponding factor can also be represented
in the form of the functional integral

\begin{equation}\label{Nielsen_Kallosh3}
\mbox{Det}\Big[\bm{\Box}^2 R\Big(\frac{\bm{\Box}}{\Lambda^2}\Big)\Big] = \int D\chi^{(+4)} D\sigma\,\exp\Big\{\frac{2i}{e_0^2}\,\mbox{tr}\int d\zeta^{(-4)} \chi^{(+4)}\,\bm{\Box}^2\, R\Big(\frac{\bm{\Box}}{\Lambda^2}\Big)\sigma\Big\},
\end{equation}

\noindent
where $\chi^{(+4)}$ and $\sigma$ are anticommuting analytic superfields in the adjoint representation.

According to ref. \cite{Slavnov:1977zf}, in the generating functional of a theory regularized by higher covariant derivatives
it is also necessary to insert the Pauli--Villars determinant in order to cancel the one-loop divergences which survive
after adding the higher derivative term to the classical action. Construction of the Pauli-Villars determinant
in the case under consideration is realized as follows. We introduce the commuting analytic superfield $P^{++} = e_0 P^{++A} t^A.$
To find its action, one considers the terms quadratic in $v^{++}$ in the sum of the regularized and gauge-fixing actions.
Then, one replaces $v^{++}$ by $P^{++}$ and adds the mass-like regularization parameter,

\begin{equation}\label{Action_PV}
S_{\mbox{\scriptsize PV}}\bigg|_{v^{++}=0} \equiv \int d\zeta^{(-4)}_1 d\zeta^{(-4)}_2 P^{++A}_1\, \frac{\delta^2 (S_{\mbox{\scriptsize reg}}
+ S_{\mbox{\scriptsize gf}})}{\delta v^{++A}_1 \delta v^{++B}_2}\, P^{++B}_2 \pm \frac{M^4}{2} \int d\zeta^{(-4)} (P^{++A})^2,
\end{equation}

\noindent
where the regularized action $S_{\mbox{\scriptsize reg}}$ was defined by eq. (\ref{Action_Of_Regularized_Theory}).
The resulting action for the superfield $P^{++}$ is almost the same as the part of the usual action quadratic in $v^{++}$,
but contains the mass term. Its sign is concordant with the one in eq. (\ref{Action_Of_The_Theory}) and is chosen in such a way
that the part of the action which does not contain (both background and quantum) gauge superfields takes the form

\begin{equation}
\pm \frac{1}{2} \int d\zeta^{(-4)} P^{++A} \Big[\partial^4 R\Big(\frac{\partial^2}{\Lambda^2}\Big) + M^4\Big] P^{++A}.
\end{equation}

\noindent
Therefore, all $P^{++}$ vertices will be exactly the same as the $v^{++}$ vertices, but the propagators will be different.
As will see later, the expression (\ref{Action_PV}) is sufficient for one-loop calculations.

Note that in eq. (\ref{Action_PV}) we present only the result for vanishing quantum gauge superfield $v^{++}$.
The dependence on $v^{++}$ can be restored from the requirement of canceling one-loop divergences of the two-point Green function
of the {\it quantum} gauge superfield. This Green function is not calculated in this paper, because it is not necessary
for obtaining the one-loop $\beta$-function. However, it can get essential in higher orders. The parameter $M$ (with the dimension of mass)
in the action (\ref{Action_PV}) should be proportional to the parameter $\Lambda$ in the regularized action (\ref{Action_Of_Regularized_Theory}),

\begin{equation}\label{PV_Mass}
M = a\Lambda,
\end{equation}

\noindent
where $a$ is an arbitrary  constant. In particular, it is possible to choose $a=1$. However,
an arbitrariness in the choice of both the constant $a$ and  the regulator function $R(x)$ implies
that the higher covariant derivative regularization is not uniquely defined.
The dependence of the renormalization group functions on the regularization parameters
is very similar to the scheme dependence. For the gauge $\beta$-function it becomes essential
starting from the three-loop approximation. The explicit dependence of the three-loop $\beta$-function(s)
on the parameter $a$ and the regulator functions for $4D$, ${\cal N}=1$ supersymmetric models
can be found in \cite{Kazantsev:2020kfl}.

Now it becomes possible to construct the Pauli--Villars determinant

\begin{equation}\label{PV_Determinant_Inverse}
\mbox{Det}^{-1}(PV,M) = \int DP^{++}\exp(i\,S_{\mbox{\scriptsize PV}}).
\end{equation}

\noindent
As we will see, in order to cancel the residual one-loop divergences,
one should insert the determinant defined in  (\ref{PV_Determinant_Inverse}) into the generating functional,
which then takes the form

\begin{eqnarray}\label{Z_Generating_Functional}
&& Z[\mbox{sources}] = \int Dv^{++} D\widetilde{q^+} Dq^+ Db\,Dc\, D\varphi\,\mbox{Det}^{1/2}\Big[\bm{\Box}^2 R\Big(\frac{\bm{\Box}}{\Lambda^2}\Big)\Big]\,\nonumber\\
&&\qquad\qquad\qquad\qquad
\times\,\mbox{Det}(PV,M)\,\exp\Big(iS_{\mbox{\scriptsize reg}} + iS_{\mbox{\scriptsize gf}}
+ iS_{\mbox{\scriptsize FP}} + iS_{\mbox{\scriptsize NK}} + iS_{\mbox{\scriptsize sources}} \Big).\qquad
\end{eqnarray}

\noindent
It is important that the expression for the generating functional contains the Pauli--Villars determinant to the degree one, while eq. (\ref{PV_Determinant_Inverse}) defines
the inverse of this determinant. This implies that any closed loop of the Pauli--Villars superfield
contributes an additional minus sign exactly as a loop of fermions. Nevertheless, the superfield $P^{++}$ is commuting and the minus sign comes
from powers of the Pauli--Villars determinant in eqs. (\ref{PV_Determinant_Inverse}) and (\ref{Z_Generating_Functional}).

The generating functional (\ref{Z_Generating_Functional}) also depends on the sources which are included into the term

\begin{equation}
S_{\mbox{\scriptsize sources}} = \int d\zeta^{(-4)} \Big(v^{++A} J^{(+2)A} + j^{(+3)A} q^{+A} + \widetilde{j^{(+3)A}} \widetilde{q^{+A}} + \ldots\Big),
\end{equation}

\noindent
where dots stand for possible sources for ghost superfields. The effective action is constructed according to the standard procedure as the Legendre transform of the generating
functional $W=-i\ln Z$ for the connected Green functions. It will depend on the background gauge superfield as a parameter and will be manifestly gauge invariant.
The manifest ${\cal N}=(1,0)$ supersymmetry is also guaranteed due to the use of the harmonic superspace.

\section{One-loop divergences}
\label{Section_One_loop}

\subsection{The general structure of divergent contributions}
\hspace*{\parindent}\label{Subsection_Quantum_Correction_Structure}

Let us analyze possible divergent contributions to the effective action. The simplest way to perform such an analysis is to resort to the dimensional considerations.
The dimensions of various quantities encountered in the harmonic superspace approach are

\begin{eqnarray}
&& [d^{14}z] = m^{-2},\qquad\quad\ [d\zeta^{(-4)}] = m^{-4},\qquad [D_a^+] = m^{1/2},\qquad [(D^+)^4] = m^2\,,\qquad\nonumber\\
&& [q^+] = [\widetilde{q^+}] = m^2,\quad\ \ [V^{\pm\pm}] =1, \qquad\qquad [F^{++}] = m^2, \qquad\ [e_0]=1. \label{Requir}
\end{eqnarray}

Any divergent contribution to the effective action should be invariant under the background gauge transformations, be  presentable as an integral over $d^{14}z$,
and possess the dimension $m^\alpha$ with $\alpha\le 0$. From the (background) gauge superfield we can construct two such expressions, namely,

\begin{eqnarray}\label{Invariant1}
&& I_1 \equiv \mbox{tr}\int d\zeta^{(-4)} (\bm{F}^{++})^2 = \mbox{tr}\int d^{14}z\, du\, \bm{V}^{--} \bm{F}^{++},\\
\label{Invariant2}
&& I_2 \equiv \sum\limits_{n=2}^\infty \frac{(-i)^n}{n}\, \mbox{tr} \int d^{14}z\,du_1\,du_2\ldots du_n\, \frac{\bm{V}^{++}_1 \bm{V}^{++}_2 \ldots \bm{V}^{++}_n}{(u_1^+ u_2^+) (u_2^+ u_3^+) \ldots (u_n^+ u_1^+)}.\qquad
\end{eqnarray}

\noindent
The first one has the same structure as the gauge part of the action (\ref{Action_Of_The_Theory}),
while the second expression (up to an insignificant dimensionful constant) coincides with the action of standard ${\cal N}=(1,0)$ SYM theory in the harmonic superspace formulation \cite{Zupnik:1986da}.
It is easy to see that the dimensions of the invariants (\ref{Invariant1}) and (\ref{Invariant2}) are

\begin{equation}
[I_1] = 1\,, \qquad [I_2] = m^{-2}.
\end{equation}

\noindent
Therefore, the momentum integrals which determine the coefficients in these expressions are,  respectively, logarithmically and quadratically divergent.
This remains true at any loop, because the coupling constant $e_0$ is dimensionless.

The structure of harmonic superspace allows us to make immediately some statements about the expected counterterms.
Indeed, it is impossible to construct the invariants from a hypermultiplet and a background gauge superfield that simultaneously
satisfy all the requirements formulated above (after eq. (\ref{Requir})).
For instance, the expression

\begin{equation}
I_3 \equiv \mbox{tr}\int d\zeta^{(-4)} \widetilde{q^+} \bm{\nabla}^{++} q^+
\end{equation}

\noindent
cannot be written as an integral over $d^{14}z$ of the local expression, and the expression

\begin{equation}
I_4 \equiv \mbox{tr}\int d\zeta^{(-4)} \widetilde{q^+} [\bm{F}^{++}, q^+] = \mbox{tr}\int d^{14}z\,du\, \widetilde{q^+} [\bm{V}^{--}, q^+]
\end{equation}

\noindent
has the dimension $m^2$. This implies that these structures cannot appear in the {\it divergent} part of the effective action\footnote{Nevertheless,
the second expression can appear in the one-loop divergences for the ordinary $6D$, ${\cal N}=(1,0)$ SYM theory,
thanks to the dimensionful coupling constant (see \cite{Buchbinder:2016url,Buchbinder:2017ozh} for details).}.
Similar arguments also apply to the ghost contributions. This is why in the theory under consideration the hypermultiplet and ghosts
are not renormalized at any loop. More precisely, this implies that the renormalization constants $Z_q$ and $Z_c$ defined by the equations

\begin{eqnarray}
&& q = e_0 q^A t^A,\qquad\qquad\ \, b=e_0 b^A t^A, \qquad\qquad\quad c = e_0 c^A t^A,\qquad\vphantom{\Big(}\nonumber\\
&& q_R = e (q_R)^A t^A,\qquad\quad b_R = e (b_R)^A t^A, \qquad\quad\ c_R = e (c_R)^A t^A,\nonumber\\
&& q \equiv \sqrt{Z_q Z_\alpha^{-1}} q_R \qquad\quad b c \equiv Z_c Z_\alpha^{-1} b_R c_R, \qquad\ \ \frac{1}{\alpha_0} = \frac{Z_\alpha}{\alpha}\qquad
\end{eqnarray}

\noindent
are equal to one.

It is necessary to check that the contributions proportional to $I_2$ cancel each other and to explore the charge renormalization
induced by the divergences proportional to $I_1$. This charge renormalization is encoded in the $\beta$-function

\begin{equation}\label{Beta_Renormalized}
\widetilde \beta(\alpha) \equiv \frac{d\alpha}{d\ln\mu}\bigg|_{\alpha_0=\mbox{\scriptsize const}}.
\end{equation}

\noindent
Usually it is defined in terms of the renormalized coupling constant $\alpha = e^2/4\pi$.
However, when analyzing quantum corrections in supersymmetric theories, it is more expedient to use another definition of the $\beta$-function,
namely in terms of the bare coupling constant $\alpha_0 = e_0^2/4\pi$,

\begin{equation}\label{Beta_Bare}
\beta(\alpha_0) \equiv \frac{d\alpha_0}{d\ln\Lambda}\bigg|_{\alpha=\mbox{\scriptsize const}}.
\end{equation}

\noindent
The functions (\ref{Beta_Renormalized}) and (\ref{Beta_Bare}) are in general different \cite{Kataev:2013eta}, but, modulo the formal change of the argument,
coincide with each other in the so-called HD+MSL scheme. In this scheme, a theory is regularized by the higher-order covariant derivatives
and the divergences are removed by minimal subtractions of logarithms. This prescription means that only powers of $\ln\Lambda/\mu$
(where $\mu$ is a renormalization point) are included into the renormalization constants.

The $\beta$-function can be calculated, starting from the two-point Green function of the background
gauge superfield $\bm{V}^{++}$. The corresponding contribution to the effective action can be presented in the form

\begin{equation}\label{Invariant_Charge_Definition}
\Gamma^{(2)}_{\bm{V}^{++}} = \pm \frac{1}{8\pi}\,\mbox{tr}
\int \frac{d^6p}{(2\pi)^6}\, d^4\theta^+\, du\, \bm{F}^{++}_{\mbox{\scriptsize linear}}(-p,\theta,u)\,
\bm{F}^{++}_{\mbox{\scriptsize linear}}(p,\theta,u)\, d^{-1}(\alpha_0, -p^2/\Lambda^2),
\end{equation}

\noindent
where

\begin{equation}\label{F_Linear_Definition}
\bm{F}^{++}_{\mbox{\scriptsize linear}}(z,u_1) = \int \frac{du_2}{(u_1^+ u_2^+)^2} (D_1^+)^4 \bm{V}^{++}(z,u_2)
\end{equation}

\noindent
is that part of $\bm{F}^{++}$ which is linear in the background gauge superfield.

According to the definition of the renormalized coupling $\alpha$, the function $d^{-1}$, being expressed through the renormalized coupling constant,
cannot depend on the cutoff $\Lambda$ in the limit $\Lambda\to \infty$. Therefore, the $\beta$-function in the HD+MSL scheme can be evaluated
according to the prescription

\begin{equation}\label{Beta_How_To_Calculate}
\widetilde\beta(\alpha\to\alpha_0)\Big|_{\mbox{\scriptsize HD+MSL}} = \beta(\alpha_0) = \alpha_0^2\,\lim\limits_{\Lambda\to\infty} \frac{d}{d\ln\Lambda} \Big(d^{-1}-\alpha_0^{-1}\Big)\bigg|_{\alpha=\mbox{\scriptsize const}}.
\end{equation}

\noindent
In $4D$, ${\cal N}=1$ supersymmetric theories regularized by higher covariant derivatives the analogous expression is expressed as integrals of double total derivatives
at any loop \cite{Stepanyantz:2019ihw}. In what follows we will see that for the theory (\ref{Action_Of_The_Theory}) this is also true  (at least, in the one-loop approximation).

\subsection{One-loop divergences produced by the hypermultiplet and ghosts}
\hspace*{\parindent}\label{Subsection_Hypermultiplet_And_Ghosts}

\begin{figure}[h]
\begin{picture}(0,4.5)
\put(0.5,2.5){\includegraphics[scale=0.4]{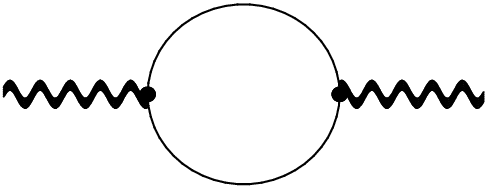}}
\put(0.5,3.7){(1)}
\put(4.5,2.5){\includegraphics[scale=0.4]{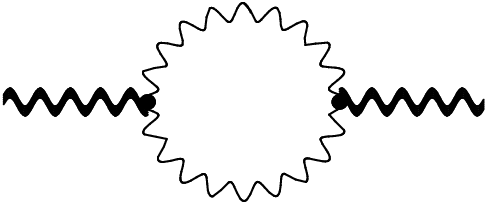}}
\put(4.5,3.7){(2)}
\put(8.5,2.5){\includegraphics[scale=0.4]{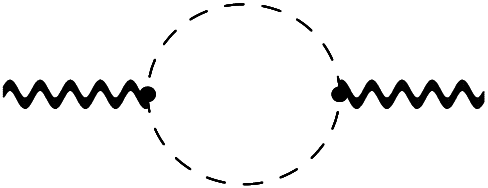}}
\put(8.5,3.7){(3)}
\put(12.5,2.5){\includegraphics[scale=0.4]{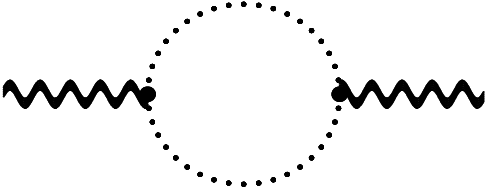}}
\put(12.5,3.7){(4)}
\put(2.5,0){\includegraphics[scale=0.4]{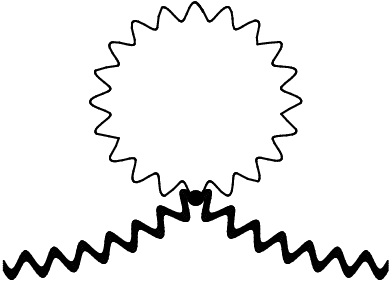}}
\put(2.3,1.6){(5)}
\put(7.0,0){\includegraphics[scale=0.4]{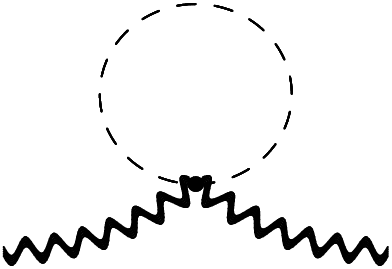}}
\put(6.8,1.6){(6)}
\put(11.5,0){\includegraphics[scale=0.4]{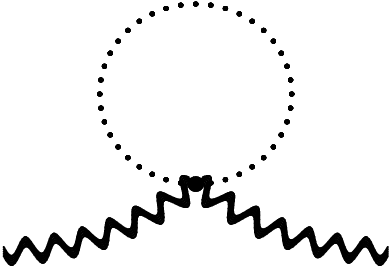}}
\put(11.3,1.6){(7)}
\end{picture}
\caption{One-loop harmonic supergraphs contributing to the two-point Green function of the background gauge superfield.}
\label{Figure_2Point_Background}
\end{figure}
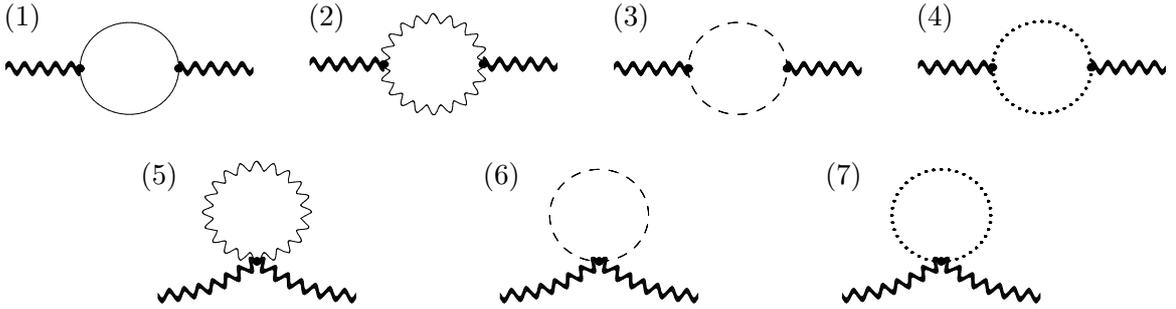

To find the coefficients of the invariants $I_1$ and $I_2$ in the divergent part
of the one-loop effective action, we should calculate the two-point Green function of the background gauge superfield in this approximation.
It is contributed by the harmonic superdiagrams presented in Fig. \ref{Figure_2Point_Background}.

In these diagrams the external bold wavy lines correspond to the background gauge superfield $\bm{V}^{++}$. Thin wavy lines denote propagators
of the quantum gauge superfield $v^{++}$ and the Pauli--Villars superfield $P^{++}$. The propagators of the Faddeev--Popov and Nielsen--Kallosh ghosts are denotes
by dashed and dotted lines, respectively. The solid lines correspond to the propagators of the hypermultiplet. From the form of the hypermultiplet action it is easy
to conclude that there are no vertices with two hypermultiplets and two external gauge lines. It should also be noticed that the diagrams with a loop of the Nielsen--Kallosh ghosts
consist of two different contributions. One of them corresponds to the superfield $\varphi$, and another is produced by the superfields $\chi^{(+4)}$ and $\sigma$, see eqs. (\ref{Nielsen_Kallosh1}) --- (\ref{Nielsen_Kallosh3}).

The hypermultiplet and ghost superdiagrams (except for the contributions of the superfields $\chi^{(+4)}$ and $\sigma$) are exactly same as in the conventional $6D$, ${\cal N}=(1,0)$ SYM theory.
According to \cite{Buchbinder:2017ozh}, their contribution to the part of the effective action producing the two-point Green function of the background gauge superfield
(in the Minkowski space, before Wick rotation) can be written as

\begin{eqnarray}\label{Hypermultiplet_And_Ghost_Contributions}
&& \Delta_{\mbox{\scriptsize ghost+hyper}}\Gamma^{(2)}_{\bm{V}^{++}} = -i C_2\, \mbox{tr} \int\frac{d^6p}{(2\pi)^6}\, d^8\theta\,\frac{du_1\,du_2}{(u_1^+ u_2^+)^2}\, \bm{V}^{++}(p,\theta,u_1) \bm{V^{++}}(-p,\theta,u_2)\qquad\nonumber\\
&& \times \int \frac{d^6k}{(2\pi)^6} \frac{1}{k^2 (k+p)^2} \Big(s_1 + s_3 + s_4 + s_6 + s_7\Big),\qquad
\end{eqnarray}

\noindent
where the quadratic Casimir $C_2$ is defined by the equation

\begin{equation}
f^{ACD} f^{BCD} = C_2 \delta^{AB},
\end{equation}

\noindent
and each numerical constant $s_n$ corresponds to the diagram $(n)$ in Fig. \ref{Figure_2Point_Background}.
The values of the constants $s_n$ based on the calculations made in \cite{Buchbinder:2017ozh} are as follows

\begin{equation}\label{Values_Of_Sn}
s_1 = 1,\qquad s_3 = -2,\qquad s_4 = 1,\qquad s_6 = s_7 = 0\,.
\end{equation}

\noindent
The contributions of the supergraphs (2) and (5) have a different structure and will be considered separately.

Note that the action for the superfields $\chi^{(+4)}$ and $\sigma$ (see eq. (\ref{Nielsen_Kallosh3})) now contains
the regulator function $R(x)$ and differs from the one considered in \cite{Buchbinder:2017ozh}. However, the contribution
of both diagrams with a loop of these superfields still vanish for exactly same reasons as in the conventional $6D$, ${\cal N}=(1,0)$ SYM theory.

Inserting the values (\ref{Values_Of_Sn}) into the expression (\ref{Hypermultiplet_And_Ghost_Contributions}),
we see that the sum of the ghost and hypermultiplet contributions vanishes,

\begin{equation}
\Delta_{\mbox{\scriptsize ghost+hyper}}\Gamma^{(2)}_{\bm{V}^{++}} = (1) + (3) + (4) + (6) + (7) = 0.
\end{equation}

\subsection{One-loop divergences produced by the quantum gauge superfield and the Pauli--Villars superfield}
\hspace*{\parindent}

In the previous section we demonstrated that the hypermultiplet and ghost contributions
to the two-point Green function of the background gauge superfield cancel each other.
Therefore, for obtaining the one-loop divergences we need to evaluate only the superdiagrams
(2) and (5) in Fig. \ref{Figure_2Point_Background}. These diagrams contain a loop of either the
quantum gauge superfield $v^{++}$ or the Pauli--Villars superfield $P^{++}$. We will calculate them,
using the analog of the Feynman gauge with $\xi_0=1$.
Details of the calculation are presented in Appendix. The result can be written in the form

\begin{eqnarray}
&&\hspace*{-5mm} \Delta_{\mbox{\scriptsize gauge}}\Gamma^{(2)}_{\bm{V}^{++}} = (2) + (5)\nonumber\\
&&\hspace*{-5mm}\quad = \mbox{tr} \int \frac{d^6p}{(2\pi)^6}\,d^4\theta^+\,du\,
\bm{F}_{\mbox{\scriptsize linear}}^{++}(-p,\theta,u) \bm{F}_{\mbox{\scriptsize linear}}^{++}(p,\theta,u) \Big(I_{v}(-p^2/\Lambda^2) - I_{P}(-p^2/\Lambda^2)\Big),\qquad
\end{eqnarray}

\noindent
where $\bm{F}_{\mbox{\scriptsize linear}}^{++A}$ defined by eq. (\ref{F_Linear_Definition})
is the part of $\bm{F}^{++A}$ linear in the background gauge superfield $\bm{V}^{++}$.
The integrals over the loop momenta $I_{v}$ and $I_{P}$ are generated by the supergraphs
containing loops of the quantum gauge superfield $v^{++}$ and the Pauli--Villars superfield $P^{++}$, respectively.
Note that (as we discussed at the end of Sect. \ref{Subsection_Quantization})
a loop of the Pauli--Villars superfield produces an additional minus sign, so that it is reasonable to put a minus sign in front of $I_P$.

It is convenient to present the result of the calculation in Appendix \ref{Appendix_Supergraph_Calculation_Background} in the form

\begin{eqnarray}
&& \frac{d}{d\ln\Lambda} I_{v} \Big|_{\Lambda\to \infty} = C_2\, \frac{d}{d\ln\Lambda} \int \frac{d^6K}{(2\pi)^6} \bigg(-\frac{4}{K^6} - \frac{R_K'}{K^4 R_K} - \frac{(R_K')^2}{K^2 R_K^2} + \frac{R_K''}{K^2 R_K}\bigg) \nonumber\\
&& = \frac{C_2}{4} \frac{d}{d\ln\Lambda} \int \frac{d^6K}{(2\pi)^6} \frac{\partial^2}{\partial K_\mu \partial K^\mu} \bigg[\frac{1}{K^4} \ln (K^4 R_K)\bigg],\\
&& \frac{d}{d\ln\Lambda} I_{P} \Big|_{\Lambda\to \infty} = C_2\, \frac{d}{d\ln\Lambda} \int \frac{d^6K}{(2\pi)^6} \bigg\{-\frac{1}{(K^4 R_K + M^4)^2}\Big(4K^2 R_K^2 \nonumber\\
&& \qquad\qquad\qquad + 4 K^4 R_K R_K' + K^6 (R_K')^2\Big) + \frac{1}{K^4 R_K+ M^4}\Big(K^2 R_K'' + 3R_K'\Big)\bigg\} \qquad\nonumber\\
&& = \frac{C_2}{4} \frac{d}{d\ln\Lambda} \int \frac{d^6K}{(2\pi)^6} \frac{\partial^2}{\partial K_\mu \partial K^\mu} \bigg[\frac{1}{K^4} \ln (K^4 R_K + M^4)\bigg],
\end{eqnarray}

\noindent
where $R_K\equiv R(K^2/\Lambda^2)$, the prime denotes the derivative with respect to $K^2$, and the loop integrals are written in the Euclidean
space, with the Wick rotation performed. The Euclidean momenta in our notation are denoted by capital letters in order
to distinguish them from the Minkowski momenta. The derivatives with respect to $\ln\Lambda$ are needed to obtain the well-defined expressions.
These derivatives should be calculated before fulfilling the integration.

Taking into account that  the function $d^{-1}$ defined by eq. (\ref{Invariant_Charge_Definition}) is equal to $\alpha_0^{-1}$ at the tree level,  in the one-loop approximation we have:

\begin{eqnarray}\label{Quantum_Corrections}
&& \frac{d}{d\ln\Lambda}(d^{-1}-\alpha_0^{-1})\Big|_{\Lambda\to\infty} = \pm 8\pi\,\frac{d}{d\ln\Lambda}(I_v - I_P)\Big|_{\Lambda\to\infty} + O(\alpha_0)\nonumber\\
&&\qquad\qquad\quad = \mp 2\pi C_2\, \frac{d}{d\ln\Lambda} \int \frac{d^6K}{(2\pi)^6} \frac{\partial^2}{\partial K_\mu\, \partial K^\mu} \bigg[\frac{1}{K^4} \ln\Big(1+ \frac{M^4}{K^4 R_K}\Big)\bigg] + O(\alpha_0).\qquad
\end{eqnarray}

\noindent
This expression is related to the $\beta$-function defined in terms of the bare coupling constant. It will be calculated in the next section.

\subsection{The one-loop $\beta$-function}
\hspace*{\parindent}

From eqs. (\ref{Beta_How_To_Calculate}) and (\ref{Quantum_Corrections}) it follows that, in the one-loop approximation, the $\beta$-function defined in terms of the bare coupling constant
for the theory (\ref{Action_Of_The_Theory}) is given by the integral of double total derivatives in the momentum space,

\begin{equation}\label{Beta_Bare_Expression}
\beta(\alpha_0) = \mp 2\pi \alpha_0^2 C_2\, \frac{d}{d\ln\Lambda} \int \frac{d^6K}{(2\pi)^6} \frac{\partial^2}{\partial K_\mu \partial K^\mu} \bigg[\frac{1}{K^4} \ln\Big(1+ \frac{M^4}{K^4 R_K}\Big)\bigg] + O(\alpha_0^3).
\end{equation}

\noindent
This expression is very similar to its $4D$ analog derived in \cite{Aleshin:2016yvj} and can be calculated by the same method. To this end,
we first perform the differentiation with respect to $\ln\Lambda$, taking into account that the parameter $M$ is proportional to $\Lambda$ (see eq. (\ref{PV_Mass})),

\begin{equation}
\beta(\alpha_0) = \pm 8\pi \alpha_0^2 C_2 \int \frac{d^6K}{(2\pi)^6} \frac{\partial^2}{\partial K_\mu \partial K^\mu}
\bigg[ \frac{M^4}{K^4(K^4 R_K + M^4)} + \frac{M^4 R_K'}{2K^2 R_K (K^4 R_K + M^4)}\bigg] + O(\alpha_0^3).
\end{equation}

\noindent
It is assumed that  the integration domain in this integral does not contain a small vicinity of the singular point $K_\mu = 0$.
In other words, we cut  a small ball of the radius $\varepsilon\to 0$ from the integration region. Using the $6D$ analog of the divergence theorem
we reduce this integral to the integral over the sphere $S^5_\infty$ with the infinitely large  radius and
the sphere $S^5_\varepsilon$ with an infinitely small radius $\varepsilon$,

\begin{equation}
\beta(\alpha_0) = \mp \frac{\alpha_0^2 C_2}{8\pi^5} \int\limits_{S^5_\infty+S^5_\varepsilon} dS_\mu^{(K)} \frac{\partial}{\partial K^\mu} \bigg[ \frac{M^4}{K^4(K^4 R_K + M^4)}+ \frac{M^4 R_K'}{2K^2 R_K (K^4 R_K + M^4)}\bigg] + O(\alpha_0^3).
\end{equation}

\noindent
The integral over the infinitely large sphere $S^5_\infty$ vanishes due to the presence of the (rapidly growing at $K^2\to\infty$) function $R_K$ in denominator.
In the remaining integral over $S^5_\varepsilon$ the only contribution that survives in the limit $\varepsilon\to 0$ is obtained if the derivative
acts on $1/K^4$ in the denominator of the first term,

\begin{equation}
\beta(\alpha_0) = \pm \frac{\alpha_0^2 C_2}{2\pi^5} \int\limits_{S^5_\varepsilon} dS_\mu^{(K)} \frac{K_\mu M^4}{K^6(K^4 R_K + M^4)} + O(\alpha_0^3).
\end{equation}

\noindent
Taking into account that the normal to the sphere $S^5_\varepsilon$ is directed inward and that the volume (understood as an analog of the surface area)
of the sphere $S^5$ with the unit radius is equal to

\begin{equation}\label{S5_Volume}
\Omega_5 = \frac{2\pi^3}{\Gamma(3)} = \pi^3,
\end{equation}

\noindent
we obtain the final expression for the one-loop $\beta$-function of the theory (\ref{Action_Of_The_Theory}) in the form:

\begin{equation}
\beta(\alpha_0) = \mp \frac{\alpha_0^2 C_2}{2\pi^2} + O(\alpha_0^3)
\end{equation}

\noindent
(the standard one-loop $\beta$-function defined in terms of the renormalized coupling constant is obtained through a formal replacement $\alpha_0\to \alpha$).
This expression agrees with the results of the calculations made in \cite{Ivanov:2005qf,Casarin:2019aqw,Buchbinder:2020tnc,Buchbinder:2020ovf},
when one takes into account the contribution of the hypermultiplet in the adjoint representation.

\section{Higher orders}
\label{Section_Higher_Orders}
\hspace*{\parindent}

The calculation of one-loop divergent quantum corrections for the theory (\ref{Action_Of_The_Theory}) demonstrated, first, the absence of quadratic divergences in this approximation
and, second,  the factorization of the integrals specifying the one-loop $\beta$-function into integrals of double total derivatives in the momentum space.
This leads us to the suggest that similar features are preserved in higher orders. If this guess is true, the absence of quadratic divergences
in all orders would imply the renormalizability of the theory.

For $4D$, ${\cal N}=1$ supersymmetric theories the factorization of integrals defining the $\beta$-function into integrals
of double total derivatives in the momentum space led  to the exact NSVZ expression for the $\beta$-function. This exact expression was obtained by summing
singular contributions which appear when double total derivatives act on the inverse-squared momenta,

\begin{equation}
\frac{\partial^2}{\partial K_\mu \partial K^\mu}\Big(\frac{1}{K^2}\Big) = -4\pi^2 \delta^4(K).
\end{equation}

\noindent
Taking eq. (\ref{S5_Volume}) into account, we find that the $6D$ analog of this equation is

\begin{equation}
\frac{\partial^2}{\partial K_\mu \partial K^\mu}\Big(\frac{1}{K^4}\Big) = -4\pi^3 \delta^6(K).
\end{equation}

Exactly as the inverse-squared momenta come from various propagators in the $4D$ case, for the $6D$ theory (\ref{Action_Of_The_Theory}) momenta to the inverse degree $-4$
arise from the gauge superfield propagators. Therefore, this theory in a certain sense is quite similar to the pure $4D$, ${\cal N}=1$ SYM theory, for which the NSVZ expression yields
the exact $\beta$-function in the form of the geometric series. This is why we are tempted to assert that the $\beta$-function for the theory (\ref{Action_Of_The_Theory})
can also be written in a similar way.

To construct the corresponding would-be exact expression, we recall how the NSVZ $\beta$-function is obtained in the $4D$ case.
If we consider a vacuum supergraph, then the sum of all superdiagrams obtained by attaching two external lines of the background gauge superfield
is given by an integral of double total derivatives. These double total derivatives effectively cut internal lines in the original vacuum supergraph
and produce the anomalous dimensions of various quantum superfields. Their sum coincides with the NSVZ equation written
in the specific form proposed in \cite{Stepanyantz:2016gtk}. Let us assume that this qualitative picture also takes place in the case under consideration,

\begin{figure}[h]
\begin{picture}(0,3.2)
\put(7.15,0.9){\includegraphics[scale=0.3]{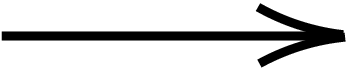}}
\put(4.3,0.05){\includegraphics[scale=0.21]{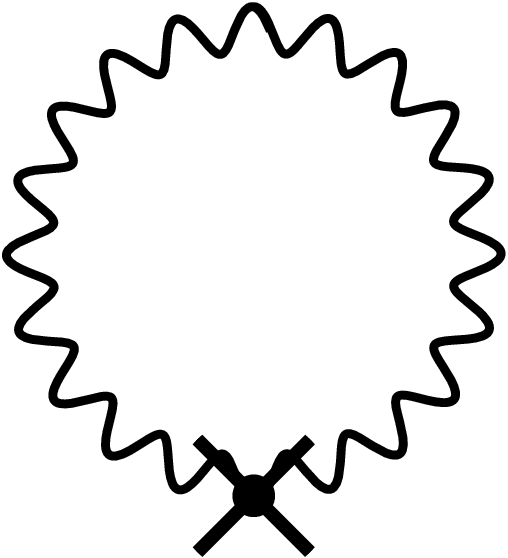}}
\put(10.0,0){\includegraphics[scale=0.21]{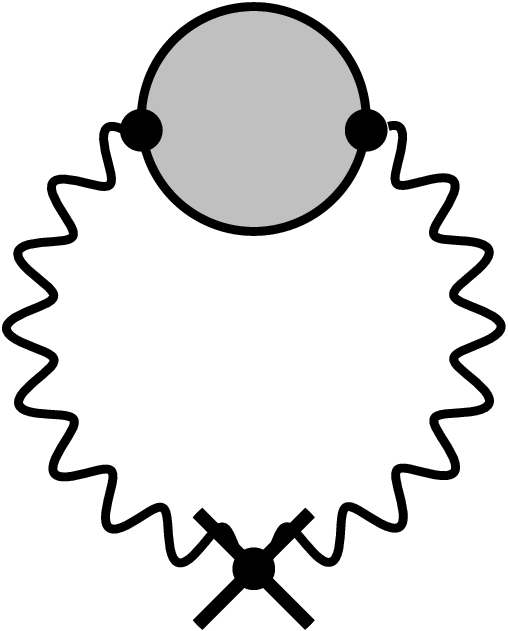}}
\end{picture}
\caption{The double total derivatives cut the gauge propagator in the one-loop approximation and the effective gauge propagator in higher orders. These cuts are denoted by the crosses.}
\label{Figure_Crosses}
\end{figure}

The cut of the one-loop supergraph is schematically presented in Fig. \ref{Figure_Crosses} on the left.
It is reasonable to suggest that in the higher orders in loops the double total derivatives cut the effective propagator, as is shown in Fig. \ref{Figure_Crosses} on the right.
The effective propagator is obtained from the two-point Green function of the quantum gauge superfield.
Due to the Slavnov--Taylor identity \cite{Taylor:1971ff,Slavnov:1972fg} the quantum corrections to this Green function are written in the form

\begin{equation}\label{Invariant_Charge_Definition_Quantum}
\Gamma^{(2)}_{v^{++}} - S^{(2)}_{\mbox{\scriptsize gf}}
= \pm \frac{1}{8\pi}\,\mbox{tr} \int \frac{d^6k}{(2\pi)^6}\, d^4\theta^+\, du\, f^{++}(-k,\theta,u)\, f^{++}(k,\theta,u)\, d_q^{-1}(\alpha_0, -k^2/\Lambda^2),
\end{equation}

\noindent
where

\begin{equation}\label{F_Quantum_Linear_Definition}
f^{++}(z,u_1) = \int \frac{du_2}{(u_1^+ u_2^+)^2} (D_1^+)^4 v^{++}(z,u_2)
\end{equation}

\noindent
is a part of $F^{++}$ linear in the quantum gauge superfield. In the tree approximation the function $d_q^{-1}$ (of the Euclidean momentum $K_\mu$ for finite values of $\Lambda$)
is equal to $\alpha_0^{-1} R(K^2/\Lambda^2)$. Therefore, this function can be written in the form

\begin{equation}\label{Pi_Definition}
d_q^{-1}(\alpha_0,K^2/\Lambda^2) = \alpha_0^{-1} R(K^2/\Lambda^2) - \alpha_0^{-1}\Pi(\alpha_0,K^2/\Lambda^2) \equiv \alpha_0^{-1}(R_K - \Pi_K),
\end{equation}

\noindent
where $\Pi$ is the polarization operator of the quantum gauge superfield.

For $4D$ supersymmetric gauge theories the renormalization of the quantum gauge superfield is related to the renormalization of the coupling constant
and the Faddeev--Popov ghosts by the relation $Z_v Z_c Z_\alpha^{-1/2}=1$, which follows from the non-renormalization theorem for the triple gauge
ghost vertices proved in \cite{Stepanyantz:2016gtk}. As we demonstrated in Sect. \ref{Subsection_Quantum_Correction_Structure}, for $6D$ theory (\ref{Action_Of_The_Theory}) the ghosts
are not renormalized, so that $Z_c=1$. Therefore, it is natural to assume that  $Z_v Z_\alpha^{-1/2}=1$ in this case, so that the quantum gauge superfields $v^{++}$
is not renormalized,

\begin{equation}
v^{++} = e_0 v^{++A} t^A = e (v_R^{++})^A t^A = v^{++}_R\equiv Z_v Z_\alpha^{-1/2} v^{++}.
\end{equation}

\noindent
If this holds, from eq. (\ref{Invariant_Charge_Definition_Quantum}) we can see that the derivative of the function $d_q$ (expressed
 in terms of the renormalized coupling constant) with respect to $\ln\Lambda$ vanishes in the limit $\Lambda\to\infty$. Hence, from eq. (\ref{Pi_Definition}) we obtain

\begin{equation}\label{Derivative_Of_R-Pi_Logarithm}
\frac{d\ln(R_K-\Pi_K)}{d\ln\Lambda}\bigg|_{\Lambda\to\infty} = \frac{d\ln(R_K-\Pi_K)}{d\ln\Lambda}\bigg|_{K\to 0} = \frac{\beta(\alpha_0)}{\alpha_0}.
\end{equation}

Looking at Fig. \ref{Figure_Crosses} and eq. (\ref{Pi_Definition}) we are tempted to suggest that the $\beta$-function
in higher orders can be obtained via replacing the function $R_K$ in the one-loop result (\ref{Beta_Bare_Expression}) b $R_K-\Pi_K$,

\begin{eqnarray}
&& \frac{\beta(\alpha_0)}{\alpha_0^2} = \pm 2\pi C_2 \int\frac{d^6K}{(2\pi)^6} \frac{d}{d\ln\Lambda} \frac{\partial^2}{\partial K_\mu \partial K^\mu}\bigg[\frac{1}{K^4}\ln \Big(K^4 (R_K-\Pi_K)\Big)
\nonumber\\
&&\qquad\qquad\qquad\qquad\qquad\qquad\qquad\qquad\qquad\qquad  - \frac{1}{K^4} \ln\Big(K^4 R_K + M^4\Big) - (PV)\bigg],\qquad
\end{eqnarray}

\noindent
where we explicitly wrote down the one-loop contribution of the Pauli--Villars superfield and denoted their contributions in higher orders by the symbol $(PV)$.
The explicit $4D$ calculations made in higher orders (see, e.g., \cite{Stepanyantz:2019lyo}) demonstrate that the Pauli--Villars superfields
are essential only in the one-loop approximation, while in higher orders they either cancel one-loop subdivergences or give vanishing contributions because
the double total derivatives act on the expressions that do not involve singularities. Presumably, exactly the same situation should take place
in the case under consideration. Then it is possible to omit the term $(PV)$ and to keep only the one-loop term containing $M$. After that,
we calculate one derivative with respect to $K_\mu$ and use the divergence theorem for the other one. As before, the integral over the infinitely
large sphere $S^5_\infty$ vanishes due to the presence of the regulator function. Therefore, the result can be written in the form

\begin{eqnarray}
&& \frac{\beta(\alpha_0)}{\alpha_0^2} = \pm \frac{C_2}{16\pi^5} \int\limits_{S^5_\varepsilon} dS_\mu^{(K)} \frac{d}{d\ln\Lambda} \bigg[\frac{K^\mu}{K^6} - \frac{2 K^\mu}{K^6}\ln \Big(K^4 (R_K-\Pi_K)\Big)
\nonumber\\
&& \qquad\qquad\qquad\qquad\qquad\qquad\qquad\qquad\qquad + \frac{2K^\mu}{K^6} \ln\Big(K^4 R_K + M^4\Big) + O\Big(\frac{1}{K^3}\Big)\bigg],\qquad
\end{eqnarray}

\noindent
where $O(1/K^3)$ denote terms which grow as $1/K^3$ or slower  in the limit $K\to 0$. Evidently, they do not contribute to the integral over
the infinitely small sphere $S^5_\varepsilon$. Next, we calculate the derivative with respect to $\ln\Lambda$ and, after that, evaluate the integral,
taking into account  eq. (\ref{S5_Volume}) and the inward direction of the normal. The result can be written as

\begin{equation}\label{Beta_Equation}
\frac{\beta(\alpha_0)}{\alpha_0^2} = \mp \frac{C_2}{2\pi^2} \pm \frac{C_2}{8\pi^2} \frac{d}{d\ln\Lambda} \ln(R_K-\Pi_K)\bigg|_{K\to 0} = \mp \frac{C_2}{2\pi^2} \pm \frac{C_2}{8\pi^2} \frac{\beta(\alpha_0)}{\alpha_0},
\end{equation}

\noindent
where we took into account eq. (\ref{Derivative_Of_R-Pi_Logarithm}). Solving eq. (\ref{Beta_Equation}) for $\beta_0$, we obtain the NSVZ-like expression
for the $\beta$-function defined in terms of the bare coupling constant,

\begin{equation}\label{Exact_Beta}
\beta(\alpha_0) = \mp \frac{\alpha_0^2 C_2}{2\pi^2\Big(1\mp \alpha_0 C_2/8\pi^2\Big)}.
\end{equation}

\noindent
The signs in this equation are concordant with the ones in eq. (\ref{Action_Of_The_Theory}). Namely,
the upper signs correspond to the upper sign in eq. (\ref{Action_Of_The_Theory}), and the lower signs
correspond to the lower sign in eq. (\ref{Action_Of_The_Theory}). Their origin can explained as follows.
Changing the sign in the gauge part of the action (\ref{Action_Of_The_Theory})
we actually change the sign in front of $\alpha_0$ and, therefore, the sign of $\beta(\alpha_0)$.

Although the reasonings used in deriving this equation look quite rigorous, the result can nevertheless be considered
merely as a guess about the form of the $\beta$-function. Actually, the structure of the one-loop quantum corrections for the theory (\ref{Action_Of_The_Theory})
is very similar to that in the $4D$ supersymmetric gauge theories. In particular, the factorization of loop integrals reducing the $\beta$-function
to integrals of double total derivatives may lead to recursive relations between the renormalization group functions, exactly as it happens in the $4D$ case
(see, e.g., \cite{Stepanyantz:2020uke}). To justify the rather lax reasoning used for deriving eq. (\ref{Exact_Beta}), we could refer to the fact
that the same speculations for the $4D$, ${\cal N}=1$ SYM theory exactly reproduced the NSVZ $\beta$-function. The two-loop verification
of eq. (\ref{Exact_Beta}) and a possible all-loop derivation of this expression seem to be interesting problems for the future study.

If the expression (\ref{Exact_Beta}) is correct and (for the $\beta$-function defined in terms of the renormalized coupling constant)
is valid in the HD+MSL scheme as in the $4D$ case, then it is possible to integrate the renormalization group equation

\begin{equation}
\frac{d\alpha}{d\ln\mu} = \widetilde\beta(\alpha) = \mp \frac{\alpha^2 C_2}{2\pi^2\Big(1\mp \alpha C_2/8\pi^2\Big)}.
\end{equation}

\noindent
It is convenient to present the result in the form of the renormalization-group invariant
(analogous to those considered in \cite{Novikov:1983uc,Kataev:2024amm,Rystsov:2024soq})

\begin{equation}
\Big(\frac{\alpha}{\mu^4}\Big)^{C_2}\exp\Big(\pm\frac{8\pi^2}{\alpha}\Big)=\mbox{RGI}.
\end{equation}

\noindent
By definition, it does not depend on the renormalization point $\mu$ in all orders.

\section{Conclusions}
\hspace*{\parindent}

In this paper we have developed an approach to studying the
structure of quantum corrections for the $6D$ theory
(\ref{Action_Of_The_Theory}) regularized by higher covariant
derivatives in the harmonic superspace. The one-loop divergences
have been calculated and analyzed in detail. The obtained results on
the structure of divergences are in agreement with the results of
the papers \cite{Buchbinder:2020tnc,Buchbinder:2020ovf}, where the
regularization by dimensional reduction has been used, and
demonstrate some new interesting features of this structure which are displayed due to
the use of the higher-derivative regularization. Certainly, when
comparing the results, one should take into account the
hypermultiplet contribution present in the theory under
consideration (\ref{Action_Of_The_Theory}).

The action of the theory under consideration already contains the fourth derivatives of the gauge superfield
and, therefore, despite we work in six-dimensions, the degree of divergence does not grow unboundedly with a number of loops.
Although the effective action could include the gauge invariant term (\ref{Invariant2}) as the potential source of the quadratically divergent contributions,
at least in the one-loop approximation no quadratic divergences appear due to the presence of the hypermultiplet in the adjoint representation. Namely, the quadratic divergences produced by the Faddeev--Popov and Nielsen--Kallosh ghosts are canceled by the quadratic divergences produced by the hypermultiplet, see eq. (\ref{Hypermultiplet_And_Ghost_Contributions}).
Stress that the absence of the residual one-loop quadratic divergences is very important for constructing the higher-derivative covariant regularization
in the case under consideration. The presence of the hypermultiplet  also leads to the cancelation of anomalies \cite{Smilga:2006ax}.
The harmonic superspace formulation is very convenient, because  in this case ${\cal N}=(1,0)$ supersymmetry is manifest at all steps of calculating quantum corrections. The choice
of the higher covariant derivative regularization is motivated by the fact that in the $4D$ case it makes manifest
some interesting features of quantum corrections. For instance, the $\beta$-function (defined in terms of the bare couplings)
in all loops proves to be given by integrals  of double total derivatives and to satisfy the NSVZ relation. This is why
it would be interesting to understand if these features are also valid in the $6D$ higher derivative theory (\ref{Action_Of_The_Theory}).
Due to the arbitrariness in the choice of the higher-derivative regulator function it becomes possible to make a nontrivial
check of the factorization of the integrals which specify the $\beta$-function in terms of integrals of the double total derivatives even in the one-loop approximation.

The calculations carried out in this paper demonstrated that the one-loop $\beta$-function
of the theory under consideration is really determined by the integrals of the double total derivatives in momentum space
for an arbitrary choice of the higher-derivative regulator function. These integrals are very similar
to those one-loop integrals which appear in $4D$ supersymmetric gauge theories. Therefore, it is natural to assert that
(exactly as in the $4D$ case) such a structure of the loop integrals also persists
in higher orders of the perturbation theory. The double total derivatives allows calculating
an  integral over one of the loop momenta, thereby reducing the number of loop integrations by one.
In $4D$ theories this gives rise to the all-loop exact $\beta$-function, and one can suppose
that an analog of this $\beta$-function also exists for the $6D$ higher-derivative theory (\ref{Action_Of_The_Theory}).

In this paper we tried to guess the form of the corresponding expression based on some non-strict indirect arguments.
Obviously, such a reasoning cannot be regarded as a proof. Nevertheless, it made possible to derive some trial expression
which can be verified (or rejected) in the future study. Definitely, it would be interesting to check it by an explicit two-loop calculation
and/or  even to prove it in all loops.

\appendix

\section*{Appendices}

\section{Propagators and vertices for the theory regularized by higher covariant derivatives}
\hspace*{\parindent}

To calculate the diagrams presented in Fig. \ref{Figure_2Point_Background}, we need the expressions for the quantum gauge superfield propagator
and vertices with two lines of the quantum gauge superfields and one or two lines of the background gauge superfield. In order to get them
we start with the action (\ref{Action_Of_Regularized_Theory}) and calculate its second variation, assuming that $\delta V^{++} = v^{++}$.
Then the terms which determine the propagator and vertices we are interested in are obtained from the expression

\begin{equation}\label{Second_Variation}
\frac{1}{2} \delta^2 S_{\mbox{\scriptsize reg}}\Big|_{V^{++}\to \bm{V}^{++},\ q^+\to 0}
\end{equation}

\noindent
(the hypermultiplet and ghost contributions to the one-loop effective action have already been presented in Sect. \ref{Subsection_Hypermultiplet_And_Ghosts}
and so we do not discuss here the corresponding Feynman rules).

The first variation of the gauge part of the action (\ref{Action_Of_Regularized_Theory}) is written as

\begin{equation}
\delta S_{\mbox{\scriptsize reg}} = \pm \frac{1}{2 e_0^2}\,\mbox{tr} \int d^{14}z\, du\,\bigg\{\, 2\,
\delta V^{--} R\Big(\frac{\Box}{\Lambda^2}\Big) F^{++} + V^{--} \delta R\Big(\frac{\Box}{\Lambda^2}\Big) F^{++}\bigg\}.
\end{equation}

\noindent
Using the identity

\begin{equation}
\delta V^{--} = \frac{1}{2} (\nabla^{--})^2 \delta V^{++} - \frac{1}{2} \nabla^{++} \nabla^{--} \delta V^{--}, \qquad \nabla^{++} F^{++} = 0,
\end{equation}

\noindent
this expression can be rewritten in the equivalent form

\begin{eqnarray}
&&\hspace*{-5mm} \delta S_{\mbox{\scriptsize reg}} = \pm \frac{1}{2e_0^2}\,\mbox{tr} \int d^{14}z\,du\,\bigg\{2\,V^{--}\,
\Box R\Big(\frac{\Box}{\Lambda^2}\Big)\delta V^{++}
+ \nabla^{--} \delta V^{--} \Big[\nabla^{++}, R\Big(\frac{\Box}{\Lambda^2}\Big) \Big] F^{++}\nonumber\\
&&\hspace*{-5mm} + V^{--} \delta R\Big(\frac{\Box}{\Lambda^2}\Big) F^{++}\bigg\}.
\end{eqnarray}

\noindent
After that, we can calculate the second variation,

\begin{eqnarray}\label{Delta2_S}
&& \delta^2 S_{\mbox{\scriptsize reg}} = \pm \frac{1}{2e_0^2}\,\mbox{tr} \int d^{14}z\,du\,\bigg\{2\, \delta V^{--}\, \Box R\Big(\frac{\Box}{\Lambda^2}\Big) \delta V^{++}
+ 2\,V^{--}\, \delta\Big[\,\Box R\Big(\frac{\Box}{\Lambda^2}\Big)\Big] \delta V^{++}
\nonumber\\
&& + \delta(\nabla^{--}) \delta V^{--} \Big[\nabla^{++}, R\Big(\frac{\Box}{\Lambda^2}\Big) \Big] F^{++}
+ \nabla^{--} \delta^2 V^{--} \Big[\nabla^{++}, R\Big(\frac{\Box}{\Lambda^2}\Big) \Big] F^{++} + \nabla^{--} \delta V^{--} \qquad\nonumber\\
&& \times \delta\Big[\nabla^{++}, R\Big(\frac{\Box}{\Lambda^2}\Big) \Big] F^{++}
+ \nabla^{--} \delta V^{--} \Big[\nabla^{++}, R\Big(\frac{\Box}{\Lambda^2}\Big) \Big] \delta F^{++} + 2\,\delta V^{--} \delta R\Big(\frac{\Box}{\Lambda^2}\Big) F^{++}
\nonumber\\
&& + V^{--} \delta^2 R\Big(\frac{\Box}{\Lambda^2}\Big) F^{++} \bigg\}.
\end{eqnarray}

\noindent
Replacing $V^{++}\to \bm{V}^{++}$, we see that the first term in this expression does not contain $\bm{F}^{++}$,
while the remaining ones are proportional either to the first or to the second powers of $\bm{F}^{++}$, because of the identities

\begin{equation}
[\nabla^{++},\Box]\, \phi^{(+2)} = i[F^{++},\phi^{(+2)}]\qquad\mbox{and}\qquad (D^+)^4 V^{--} = F^{++},
\end{equation}

\noindent
where $\phi^{(+2)}$ is an analytic superfield in the adjoint representation.

Because any perturbative contribution to the effective action can be presented
as an integral over $d^{14}z$, the terms quadratic in $\bm{F}^{++}$ correspond to the convergent loop integrals
and do not affect the divergent part of the effective action, see Sect. \ref{Subsection_Quantum_Correction_Structure} for the detailed discussion.
Therefore, in what follows they can be omitted.

Next, we should calculate the expression (\ref{Second_Variation}) and add the gauge-fixing action (\ref{Gauge_Fixing_Term}).
Integrating by parts with respect to the harmonic derivative $\bm{\nabla}_2^{++}$ and using the identities

\begin{equation}
e^{-i\bm{b}} \bm{\nabla}^{++} e^{i\bm{b}} = D^{++},\qquad D_1^{++} \frac{(u_1^- u_2^+)}{(u_1^+ u_2^+)^3} = \frac{1}{(u_1^+ u_2^+)^2} - \frac{1}{2} (D_1^{--})^2 \delta^{(2,-2)}(u_1,u_2),
\end{equation}

\noindent
it can be rewritten in the form

\begin{eqnarray}\label{Gauge_Fixing_Rewritten}
&&\hspace*{-5mm} S_{\mbox{\scriptsize gf}} = \mp \frac{1}{2 e_0^2 \xi_0} \mbox{tr} \int d^{14}z\,du_1\,du_2\, e^{i\bm{b}_1} e^{-i\bm{b}_2} v_2^{++} e^{i\bm{b}_2} e^{-i\bm{b}_1} \bigg\{\Big(\frac{1}{(u_1^+ u_2^+)^2} - \frac{1}{2} (D_1^{--})^2 \delta^{(2,-2)}(u_1,u_2)\Big)\nonumber\\
&&\hspace*{-5mm} \times \bm{\Box}_1 R\Big(\frac{\bm{\Box}_1}{\Lambda^2}\Big) v_1^{++}
+ \frac{(u_1^- u_2^+)}{(u_1^+ u_2^+)^3} \Big[\bm{\nabla}_1^{++}, \bm{\Box}_1 R\Big(\frac{\bm{\Box}_1}{\Lambda^2}\Big)\Big] v_1^{++}\bigg\}.
\end{eqnarray}

\noindent
Taking into account that

\begin{equation}
\delta V_1^{--} = \int \frac{du_2}{(u_1^+ u_2^+)^2} e^{i\bm{b}_1} e^{-i\bm{b}_2} \delta V_2^{++} e^{i\bm{b}_2} e^{-i\bm{b}_1}
= \int \frac{du_2}{(u_1^+ u_2^+)^2} e^{i\bm{b}_1} e^{-i\bm{b}_2} v_2^{++} e^{i\bm{b}_2} e^{-i\bm{b}_1}\,,
\end{equation}

\noindent
we see that the sum of such terms in eqs. (\ref{Delta2_S}) and (\ref{Gauge_Fixing_Rewritten}) which do not contain $\bm{F}^{++}$
in the Feynman gauge $\xi_0=1$ can be written in the form

\begin{equation}\label{Zero_Term}
\frac{1}{2} \delta^2 S_{\mbox{\scriptsize reg}}\Big|_{V^{++}\to \bm{V}^{++},\ q^+\to 0} + S_{\mbox{\scriptsize gf}}\Big|_{\xi_0=1}
= \pm \frac{1}{2e_0^2}\, \mbox{tr} \int d\zeta^{(-4)} v^{++} \bm{\Box}^2 R\Big(\frac{\bm{\Box}}{\Lambda^2}\Big) v^{++} + \ldots\,,
\end{equation}

\noindent
where dots stand for terms which include $\bm{F}^{++}$. The term that does not depend on the background gauge superfield
determines the propagator of the quantum gauge superfield

\begin{equation}
\frac{\delta^2 \ln Z_0}{\delta J^{(+2)A}_1 \delta J^{(+2)B}_2} = \mp \frac{2i}{\partial^4 R} (D_1^+)^4 \delta^{14}(z_1-z_2)\,\delta^{(-2,2)}(u_1,u_2)\, \delta^{AB}.
\end{equation}

\noindent
However, the expression (\ref{Zero_Term}) also contributes to the vertices. These vertices are essential, when calculating the one-loop divergences.
In particular, after some calculations it is possible to obtain that the vertex containing two legs of the quantum gauge superfield and
one leg of the background gauge superfield can be presented in the form

\begin{eqnarray}\label{Vertex_1}
&& \mbox{Ver}_1 = \pm \frac{e_0}{8} f^{ABC} \int \frac{d^6p}{(2\pi)^6} \frac{d^6k}{(2\pi)^6} d^{8}\theta\, du\, \frac{(k+p)^4 R_{k+p} - k^4 R_k}{(k+p)^2-k^2}\, v^{++A}(-k-p,\theta,u)
\qquad\nonumber\\
&& \times \bigg[2\bm{V}^{--B}(p,\theta,u)\,D^{--} v^{++C}(k,\theta,u) + D^{--} \bm{V}^{--B}(p,\theta,u)\, v^{++C}(k,\theta,u)\bigg],\qquad
\end{eqnarray}

\noindent
where $R_k\equiv R(-k^2/\Lambda^2)$. Strictly speaking, this expression also contains a term quadratic in $\bm{V}^{++}$, see eq. (\ref{V--}).  However,
it is easy to see that, because of the total antisymmetry of the structure constants, these quadratic terms do not contribute to the tadpole
diagram (5) in Fig. \ref{Figure_2Point_Background}. In what follows we will always assume that only the part linear
in $\bm{V}^{++}$ is taken from $\bm{V}^{--}$ in the expression (\ref{Vertex_1}).

Similarly, it is possible to evaluate that contribution to the vertex quadratic in the background gauge superfield which comes from the expression (\ref{Zero_Term}).
In this case the calculations are more complicated and the result is given by the rather bulky expression

\begin{eqnarray}\label{Vertex_2}
&&\hspace*{-5mm} \mbox{Ver}_2 = \mp \frac{e_0^2}{16} f^{ABC} f^{CDE} \int \frac{d^6p}{(2\pi)^6} \frac{d^6q}{(2\pi)^6} \frac{d^6k}{(2\pi)^6}\,  d^{8}\theta\, du\, v^{++A}(-k-q-p,\theta,u) \nonumber\\
&&\hspace*{-5mm} \times \bigg\{
\frac{2((k+q+p)^4 R_{k+q+p} - q^4 R_q)}{(k+q+p)^2-q^2}\, \bm{V}^{--B}(k,\theta,u)\,\bm{V}^{--D}(p,\theta,u)\, v^{++E}(q,\theta,u)\nonumber\\
&&\hspace*{-5mm} - \bigg[\frac{(q+p)^4 R_{q+p}}{((q+p)^2-(k+q+p)^2)((q+p)^2-q^2)} + \frac{q^4 R_{q}}{(q^2-(k+q+p)^2)(q^2-(q+p)^2)}
\nonumber\\
&&\hspace*{-5mm} + \frac{(k+q+p)^4 R_{k+q+p}}{((k+q+p)^2-(q+p)^2)((k+q+p)^2-q^2)} \bigg]
\Big(2\bm{V}^{--B}(k,\theta,u) D^{--} \nonumber\\
&&\hspace*{-5mm} + D^{--}\bm{V}^{--B}(k,\theta,u) \Big) (D^+)^4 \Big(2\bm{V}^{--D}(p,\theta,u) D^{--} + D^{--}\bm{V}^{--D}(p,\theta,u)\Big)\nonumber\\
&&\hspace*{-5mm} \times v^{++E}(q,\theta,u)\bigg\}.
\end{eqnarray}

\noindent
Strictly speaking, this expression is not quadratic in $\bm{V}^{++}$ because each $\bm{V}^{--}$ is an infinite series in $\bm{V}^{++}$. For constructing
the vertex with two external lines of $\bm{V}^{++}$ we need to take only linear parts from each $\bm{V}^{--}$. It should be also noted that the terms quadratic
in the background gauge superfield which are present in eq. (\ref{Vertex_1}) were not included into eq. (\ref{Vertex_2}).

We will also need the terms proportional to the first power of $\bm{F}^{++}$. Some of them come from the expression (\ref{Delta2_S}), namely, from the terms

\begin{eqnarray}\label{Contribution1}
&& \frac{1}{2}\delta^2 S_{\mbox{\scriptsize reg}} \to \pm \frac{1}{4e_0^2}\,\mbox{tr} \int d^{14}z\,du\,\bigg\{ 2\,V^{--}\,
\delta\Big[\,\Box R\Big(\frac{\Box}{\Lambda^2}\Big)\Big] \delta V^{++} + 2\,\delta V^{--} \delta R\Big(\frac{\Box}{\Lambda^2}\Big) F^{++} \qquad\nonumber\\
&& + \nabla^{--} \delta V^{--} \delta\Big[\nabla^{++}, R\Big(\frac{\Box}{\Lambda^2}\Big) \Big] F^{++} +
\nabla^{--} \delta V^{--} \Big[\nabla^{++}, R\Big(\frac{\Box}{\Lambda^2}\Big) \Big] \delta F^{++} \bigg\}
\end{eqnarray}

\noindent
after the replacement $\delta V^{++} \to v^{++}$ and $V^{++}\to \bm{V}^{++}$.

The terms proportional to $\bm{F}^{++}$ also come from the gauge-fixing action (\ref{Gauge_Fixing_Term}).
In the Feynman gauge the contribution of the considered structure reads

\begin{equation}\label{Contribution2}
S_{\mbox{\scriptsize gf}}\Big|_{\xi_0=1} \to \mp \frac{1}{2e_0^2} \mbox{tr} \int d^{14}z\,du_1\,du_2\, e^{i\bm{b}_1} e^{-i\bm{b}_2} v_2^{++} e^{i\bm{b}_2} e^{-i\bm{b}_1} \frac{(u_1^- u_2^+)}{(u_1^+ u_2^+)^3} \Big[\bm{\nabla}_1^{++}, \bm{\Box}_1 R\Big(\frac{\bm{\Box}_1}{\Lambda^2}\Big)\Big] v_1^{++}.
\end{equation}

\noindent
Starting from the expressions (\ref{Contribution1}) and (\ref{Contribution2}), after some calculations we obtain the vertex with two legs of the quantum
gauge superfield and one leg of the background gauge superfield corresponding to $\bm{F}^{++}$. The result can be written in the form

\begin{eqnarray}\label{Vertex_3}
&&\hspace*{-5mm} \mbox{Ver}_3 = \pm \frac{e_0}{8} f^{ABC} \int \frac{d^6p}{(2\pi)^6}\, \frac{d^6k}{(2\pi)^6}\, d^8\theta\, du_1\,du_2\,\bigg\{\frac{2(u_1^- u_2^+)}{(u_1^+ u_2^+)^3}\, v^{++A}(-p-k,\theta,u_2) \bm{F}^{++B}(p,\theta,u_1)\nonumber\\
&&\hspace*{-5mm}  \times\, v^{++C}(k,\theta,u_1)\, \frac{(k+p)^2 R_{k+p}- k^2 R_k}{(k+p)^2-k^2} - \frac{1}{(u_1^+ u_2^+)^2}\, v^{++A}(-p-k,\theta,u_2)\bigg[\,D_1^{--} \bm{F}^{++B}(p,\theta,u_1)\nonumber\\
&&\hspace*{-5mm}  \times\, v^{++C}(k,\theta,u_1) - \bm{F}^{++B}(p,\theta,u_1) D_1^{--} v^{++C}(k,\theta,u_1) \bigg]  \frac{k^2 R_k- p^2 R_p}{k^2-p^2}
+ \int \frac{du_3}{(u_1^+ u_2^+)^2 (u_1^+ u_3^+)^2}  \nonumber\\
&&\hspace*{-5mm} \times\, \frac{R_{k+p}-R_p}{(k+p)^2-p^2}\, \bigg[ \Big(D_1^{--} - \frac{2(u_1^- u_2^+)}{(u_1^+ u_2^+)}\Big)(D_1^+)^4\, v^{++A}(-p-k,\theta,u_2)\Big)\,\bm{F}^{++B}(p,\theta,u_1)  \nonumber\\
&&\hspace*{-5mm} - (D_1^+)^4\, v^{++A}(-p-k,\theta,u_2)\, D_1^{--} \bm{F}^{++B}(p,\theta,u_1) \bigg] v^{++C}(k,\theta,u_3)\, + 2\int \frac{du_3\,(u_1^- u_2^+)}{(u_1^+ u_2^+)^3 (u_1^+ u_3^+)^2} \nonumber\\
&&\hspace*{-5mm} \times\, \bigg[\frac{R_{k+p}-R_p}{(k+p)^2-p^2} - \frac{R_{k+p}-R_k}{(k+p)^2-k^2}\bigg]\, v^{++A}(-p-k,\theta,u_2)
\bm{F}^{++B}(p,\theta,u_1) (D_1^+)^4\, v^{++C}(k,\theta,u_3) \bigg\}.\nonumber\\
\end{eqnarray}

The contribution proportional to $\bm{F}^{++}$ in the vertex with two legs of the quantum gauge superfield and two legs of the background gauge superfield is rather bulky and complicated.
That is why we do not present here the general expression for it. Below we write down only those terms which nontrivially contribute
to the divergent part of the one-loop effective action. In particular, in the expression (\ref{Vertex_4})
below we do not include the following parts:

1. The terms that contain less than four spinor derivatives distributed among the quantum gauge superfields.
They do not contribute to the supergraph (5) in Fig. \ref{Figure_2Point_Background} because of the identity

\begin{equation}
(D^+)^\alpha \delta^8(\theta_1-\theta_2)\Big|_{\theta_1=\theta_2} = 0\qquad\mbox{for any}\quad \alpha\le 7
\end{equation}

\noindent
(four spinor derivatives come from the gauge propagator).

2. The terms containing more than four spinor derivatives acting on the external legs, which produce loop integrals convergent
in the ultraviolet region. Note that $\bm{F}^{++}$ already contains four spinor derivatives acting on the background gauge superfield.

3. The terms in which the number of the harmonic derivatives $D^{--}$ is so small that the corresponding contributions
to the tadpole graph vanish due to the identity $(u^+ u^+) = 0$.

4. The terms for which the corresponding contributions to the tadpole graph vanish due to the identities $f^{AAB} = 0$ or $(\bm{b}_1-\bm{b}_2)\Big|_{u_1=u_2}=0$.

Moreover, as we will see in the next section, for calculating the contribution of the logarithmically divergent part of the supergraph (5)
in Fig. \ref{Figure_2Point_Background} to the $\beta$-function, we need to consider a limit of the vanishing external momentum. In this case the background gauge superfields
have zero momenta, and the momenta of the quantum gauge superfields differ only in sign. Therefore, for calculating the logarithmically
divergent part of the supergraph (5) it is sufficient to take the vertex
in the limit of the vanishing external momenta. Namely, we write the vertex under consideration in the form

\begin{equation}
\mbox{Ver}_4 = \int \frac{d^6p}{(2\pi)^6}\,\frac{d^6q}{(2\pi)^6}\,\frac{d^6k}{(2\pi)^6}\,\mbox{Ver}_4(p,q,k),
\end{equation}

\noindent
where $p$ and $q$ denote the momenta of the background gauge superfield. The momentum of one quantum gauge superfield is denoted by $k$, and the second quantum gauge superfield
has the momentum $-q-k-p$. Then (after some transformations) the vertex under consideration can be cast in the form

\begin{eqnarray}\label{Vertex_4}
&& \mbox{Ver}_4(p=0,q=0,k) = \pm \frac{e_0^2}{4} f^{ABC} f^{CDE} \int d^8\theta\, du_1\, du_2\, \frac{(u_1^- u_2^+)}{(u_1^+ u_2^+)^3} \,(D_1^+)^4 v^{++A}(-k,\theta,u_2)
\qquad\nonumber\\
&& \times\, \bm{F}_{\mbox{\scriptsize linear}}^{++B}(0,\theta,u_1) \bm{V}_{\mbox{\scriptsize linear}}^{--D}(0,\theta,u_1) \Big((k^2 R_k)'' + R_k'\Big) D_1^{--} v^{++E}(k,\theta,u_1)\nonumber\\
&&\qquad\qquad\qquad\qquad\qquad +\  \mbox{the terms that do not contribute to the supergraph (5)},\vphantom{\frac{1}{2}}
\end{eqnarray}

\noindent
where primes denote derivatives with respect to $k^2$. The terms collected in eq. (\ref{Vertex_4}) are sufficient to calculate the supergraph (5)
in Fig. \ref{Figure_2Point_Background}. We briefly describe this calculation in the next section.

\section{Details of the harmonic supergraph calculation}
\hspace*{\parindent}\label{Appendix_Supergraph_Calculation_Background}

\begin{figure}[h]
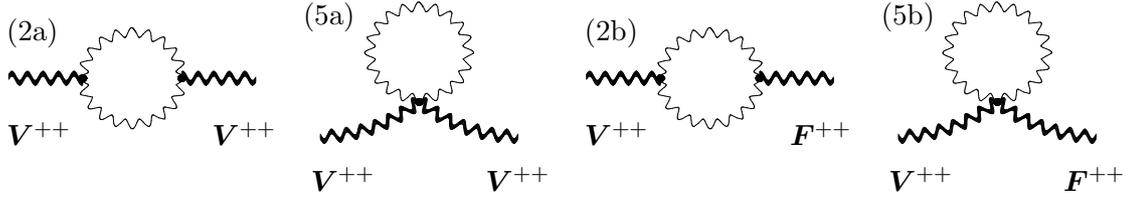

\begin{picture}(0,3.0)
\put(0.8,0.8){\includegraphics[scale=0.4]{1loop_gauge1.eps}}
\put(0.8,2.0){(2a)} \put(0.8,0.6){$\bm{V}^{++}$}  \put(3.5,0.6){$\bm{V}^{++}$}
\put(4.9,0.6){\includegraphics[scale=0.4]{1loop_gauge2.eps}}
\put(4.7,2.2){(5a)} \put(4.8,0){$\bm{V}^{++}$}  \put(7.1,0){$\bm{V}^{++}$}
\put(8.4,0.8){\includegraphics[scale=0.4]{1loop_gauge1.eps}}
\put(8.4,2.0){(2b)} \put(8.4,0.6){$\bm{V}^{++}$}  \put(11.1,0.6){$\bm{F}^{++}$}
\put(12.5,0.6){\includegraphics[scale=0.4]{1loop_gauge2.eps}}
\put(12.3,2.2){(5b)} \put(12.4,0){$\bm{V}^{++}$}  \put(14.7,0){$\bm{F}^{++}$}
\end{picture}
\caption{Various contributions to the two-point Green function of the background gauge superfield from the harmonic supergraphs containing a loop of the quantum gauge superfield.}
\label{Figure_2Point_Subdiagrams}
\end{figure}

With the help of the vertices constructed in the previous section we can calculate the divergent contributions
to the one-loop effective action produced by the superdiagrams containing a loop of the quantum gauge superfield. These superdiagrams
appear in Fig. \ref{Figure_2Point_Background} with numbers (2) and (5). It is expedient to split them into parts containing vertices
proportional to various powers of $\bm{F}^{++}$. In Sect. \ref{Subsection_Quantum_Correction_Structure} we demonstrated that the terms containing
two (or more) $\bm{F}^{++}$ correspond to finite contributions (because the result of the calculation should always be given by an integral over $d^{14}z\, du$).
Therefore, the diagrams containing these terms can be omitted. The remaining supergraphs are presented in Fig. \ref{Figure_2Point_Subdiagrams}.
They are either do not contain $\bm{F}^{++}$ at all or are linear in it. The former superdiagrams are quadratically divergent, while the latter ones diverge logarithmically.

In the quadratically divergent graphs (2a) and (5a) the vertices are generated by the expression (\ref{Zero_Term}). They are given by eqs. (\ref{Vertex_1}) and (\ref{Vertex_2}).
Note that, calculating these diagrams, it is impossible to set the external momentum $p_\mu$ equal to 0, because the sub-leading terms proportional to $p^2$ are logarithmically divergent.
However, in fact, the contributions (2a) and (5a) vanish. Namely, let us first consider the superdiagram (2a). It is given by the expression

\begin{eqnarray}
&&\hspace*{-11mm} - \frac{iC_2}{4} \mbox{tr} \int \frac{d^6p}{(2\pi)^6}\, \frac{d^6k}{(2\pi)^6}\, d^8\theta_1\,d^8\theta_2\, \frac{du_1\,du_2\,du_3\,du_4}{(u_1^+ u_3^+)^2 (u_2^+ u_4^+)^2}
\bm{V}^{++}(-p,\theta_1,u_3)\,\bm{V}^{++}(p,\theta_2,u_4) \nonumber\\
&&\hspace*{-11mm} \times\, \frac{((k+p)^4 R_{k+p} - k^4 R_k)^2}{k^4 R_k (k+p)^4 R_{k+p} ((k+p)^2-k^2)^2}
\bigg[\Big(D_1^{--} - \frac{(u_1^- u_3^+)}{(u_1^+ u_3^+)}\Big) \Big(D_2^{--} - \frac{(u_2^- u_4^+)}{(u_2^+ u_4^+)}\Big) (D^+_1)^4 \nonumber\\
&&\hspace*{-11mm} \times\, \delta^8(\theta_1-\theta_2) \delta^{(-2,2)}(u_1,u_2)\cdot (D^+_2)^4 \delta^8(\theta_1-\theta_2) \delta^{(-2,2)}(u_2,u_1)
- \Big(D_1^{--} - \frac{(u_1^- u_3^+)}{(u_1^+ u_3^+)}\Big)\nonumber\\
&&\hspace*{-11mm} \times\, (D^+_1)^4 \delta^8(\theta_1-\theta_2) \delta^{(-2,2)}(u_1,u_2)\cdot
\Big(D_2^{--} - \frac{(u_2^- u_4^+)}{(u_2^+ u_4^+)}\Big) (D^+_2)^4 \delta^8(\theta_1-\theta_2) \delta^{(-2,2)}(u_2,u_1)
\bigg].
\end{eqnarray}

\noindent
First, we note that the minimal number of spinor covariant derivatives placed between two $\delta^8(\theta_1-\theta_2)$ is equal to 8. Therefore,
 after integrating by parts with respect to $(D_2^+)^4$ all derivatives will act on $(D_1^+)^4 \delta^8(\theta_1 - \theta_2)$. After that, we use the identity

\begin{equation}\label{8_Derivatives}
\delta^8(\theta_1-\theta_2) (D_1^+)^4 (D_2^+)^4 \delta^8(\theta_1-\theta_2) = (u_1^+ u_2^+)^4 \delta^8(\theta_1-\theta_2).
\end{equation}

\noindent
Next, we note that no more than two harmonic derivatives $D^{--}$ can act on this structure. Therefore, the result for the supergraph vanishes because

\begin{equation}\label{Vanishing_Harmonics}
(u_1^+ u_2^+)\Big|_{u_1=u_2} = 0.
\end{equation}

The second quadratically divergent supergraph (5a) vanishes for the same reason. Indeed, various terms in the vertex (\ref{Vertex_2}) contain no more
than two harmonic derivatives. Therefore, they cannot produce a non-vanishing result when acting on the expression

\begin{equation}
(u_1^+ u_2^+)^4,
\end{equation}

\noindent
which comes from the gauge propagator, after setting $u_2=u_1$ (for this purpose one would need at least 4 derivatives $D^{--}$). Thus, the graph (5a) also gives the zero contribution.

The contributions of the supergraphs (2a) and (5a) with a loop of the Pauli--Villars superfield vanish for the same reasons.
The expressions for them are obtained by replacing the propagators $(k^4 R_k)^{-1}$ by $(k^4 R_k+M^4)^{-1}$ and changing the overall sign of a supergraph.

Taking into account that the supergraphs (2a) and (5a) do not contribute to the one-loop effective action, while the ghosts and hypermultiplet contributions
cancel each other, we conclude that there are no quadratic one-loop divergences in the considered theory.

The supergraphs (2b) and (5b) are logarithmically divergent. It is these supergraphs that determine the one-loop $\beta$-function of the theory (\ref{Action_Of_The_Theory}).
We calculate the $\beta$-function according to eq. (\ref{Beta_How_To_Calculate}), taking into account that the limit $\Lambda\to \infty$ is equivalent
to the limit of the vanishing external momentum $p\to 0$. In what follows, for each supergraph we will present the corresponding contribution to the expression

\begin{eqnarray}\label{Expression_For_Calculating}
&&\hspace*{-5mm} \frac{d}{d\ln\Lambda} \Delta\Gamma^{(2)}_{\bm{V}^{++},\infty} = \pm \frac{1}{8\pi}\mbox{tr} \int \frac{d^6p}{(2\pi)^6}\, d^8\theta\, du\, \bm{V}^{--}_{\mbox{\scriptsize linear}} \bm{F}^{++}_{\mbox{\scriptsize linear}}
\,\frac{d}{d\ln\Lambda} \Big(d^{-1}(\alpha_0,-p^2/\Lambda^2) - \alpha_0^{-1}\Big)\bigg|_{\Lambda \to \infty} \nonumber\\
&&\hspace*{-5mm} = \pm \frac{1}{8\pi} \frac{d}{d\ln\Lambda} \Big(d^{-1}(\alpha_0,-p^2/\Lambda^2) - \alpha_0^{-1}\Big)\bigg|_{p\to 0} \mbox{tr}\int \frac{d^6p}{(2\pi)^6}\, d^8\theta\, du\, \bm{V}^{--}_{\mbox{\scriptsize linear}}(-p,\theta,u) \bm{F}^{++}_{\mbox{\scriptsize linear}}(p,\theta,u)\nonumber\\
&&\hspace*{-5mm} = \pm \frac{1}{8\pi}\cdot \frac{\beta(\alpha_0)}{\alpha_0^2}\,\mbox{tr} \int d\zeta^{(-4)} (\bm{F}^{++}_{\mbox{\scriptsize linear}})^2,
\end{eqnarray}

\noindent
where $\Delta\Gamma$ is a sum of quantum corrections to the effective action. The subscript $\bm{V}^{++}$ and the superscript $(2)$ mean that we consider only the terms corresponding to the two-point Green function of the background gauge superfield, and the subscript $\infty$ denotes their divergent part. The derivative with respect to $\ln\Lambda$ should be taken at a fixed value of the {\it renormalized} coupling constant $\alpha$.

Let us start with the supergraph (2b) in Fig. \ref{Figure_2Point_Subdiagrams}. To avoid too large expressions, we describe in details only the contribution
produced by the first term in the expression (\ref{Vertex_3}) for the right vertex in the case when the loop corresponds to the quantum
gauge superfield $v^{++}$\footnote{The contribution of the supergraph with the Pauli--Villars superfield will be added in what follows.}.
The Feynman rules give for it the expression

\begin{eqnarray}\label{2B_First_Term_Contribution}
&&\hspace*{-6mm} -\frac{ie_0^2C_2}{8} \frac{d}{d\ln\Lambda} \int \frac{d^{6}p}{(2\pi)^6} \frac{d^6k}{(2\pi)^6} \int d^8\theta_1\,d^8\theta_2 \int du_1\, du_2\, du_3\, \frac{(u_2^- u_3^+)}{(u_2^+ u_3^+)^3}\,
\frac{1}{k^4 R_k (k+p)^4 R_{k+p}} \nonumber\\
&&\hspace*{-6mm} \times \frac{(k+p)^4 R_{k+p}- k^4 R_k}{(k+p)^2-k^2}\cdot \frac{(k+p)^2 R_{k+p} - k^2 R_k}{(k+p)^2-k^2}\, \bm{F}_{\mbox{\scriptsize linear}}^{++A}(p,\theta_2,u_2)\,
\bigg\{ \Big(2\bm{V}_{\mbox{\scriptsize linear}}^{--A}(-p,\theta_1,u_1) D_1^{--} \nonumber\\
&&\hspace*{-6mm} + D_1^{--} \bm{V}_{\mbox{\scriptsize linear}}^{--A}(-p,\theta_1,u_1)\Big) \Big((D_1^+)^4 \delta^8(\theta_1-\theta_2) \delta^{(-2,2)}(u_1,u_2)\Big) (D_3^+)^4 \delta^8(\theta_1-\theta_2) \delta^{(2,-2)}(u_1,u_3)
\nonumber\\
&&\hspace*{-6mm} - \Big(2\bm{V}_{\mbox{\scriptsize linear}}^{--A}(-p,\theta_1,u_1) D_1^{--} + D_1^{--} \bm{V}_{\mbox{\scriptsize linear}}^{--A}(-p,\theta_1,u_1)\Big)\, \Big((D_1^+)^4 \delta^8(\theta_1-\theta_2) \delta^{(-2,2)}(u_1,u_3)\Big)\nonumber\\
&&\hspace*{-7mm} \times (D_2^+)^4 \delta^8(\theta_1-\theta_2) \delta^{(2,-2)}(u_1,u_2) \bigg\}.
\end{eqnarray}

\noindent
Here the propagators give $(k^4 R_k)^{-1}$ and $((k+p)^4 R_{k+p})^{-1}$, where $R_k \equiv R(-k^2/\Lambda^2)$, and the other functions of the momenta are produced by the vertices.

The products of the Grassmannian $\delta$-functions do not vanish only if there are at least 8 spinor derivatives $D^+$ acting on them. Then, using the identity (\ref{8_Derivatives})
and doing the integral over $\theta_2$, the expression (\ref{2B_First_Term_Contribution}) can be cast in the form

\begin{eqnarray}
&&\hspace*{-6mm} -\frac{ie_0^2C_2}{8} \frac{d}{d\ln\Lambda} \int \frac{d^{6}p}{(2\pi)^6} \frac{d^6k}{(2\pi)^6} \int d^8\theta \int du_1\, du_2\, du_3\, \frac{(u_2^- u_3^+)}{(u_2^+ u_3^+)^3}\,
\frac{1}{k^4 R_k (k+p)^4 R_{k+p}}  \nonumber\\
&&\hspace*{-6mm} \times \frac{(k+p)^4 R_{k+p}- k^4 R_k}{(k+p)^2-k^2}\cdot \frac{(k+p)^2 R_{k+p} - k^2 R_k}{(k+p)^2-k^2}\, \bm{F}_{\mbox{\scriptsize linear}}^{++A}(p,\theta,u_2)\,
\bigg\{ \Big(2\bm{V}_{\mbox{\scriptsize linear}}^{--A}(-p,\theta,u_1) D_1^{--}\nonumber\\
&&\hspace*{-6mm} + D_1^{--} \bm{V}_{\mbox{\scriptsize linear}}^{--A}(-p,\theta,u_1)\Big) \Big((u_1^+ u_3^+)^4 \delta^{(-2,2)}(u_1,u_2)\Big) \delta^{(2,-2)}(u_1,u_3)
- \Big(2\bm{V}_{\mbox{\scriptsize linear}}^{--A}(-p,\theta,u_1) D_1^{--}
\nonumber\\
&&\hspace*{-6mm} + D_1^{--} \bm{V}_{\mbox{\scriptsize linear}}^{--A}(-p,\theta,u_1)\Big) \Big((u_1^+ u_2^+)^4 \delta^{(-2,2)}(u_1,u_3)\Big) \delta^{(2,-2)}(u_1,u_2)
\bigg\}.
\end{eqnarray}

\noindent
Applying eq. (\ref{Vanishing_Harmonics}) and using the relation

\begin{equation}
(u_1^+ u_3^+)^4 \delta^{(-2,2)}(u_1,u_2) = (u_2^+ u_3^+)^4 \delta^{(2,-2)}(u_1,u_2),
\end{equation}

\noindent
this contribution can be written as

\begin{eqnarray}
&& -\frac{ie_0^2C_2}{4} \frac{d}{d\ln\Lambda} \int \frac{d^{6}p}{(2\pi)^6} \frac{d^6k}{(2\pi)^6} \int d^8\theta \int du_1\, du_2\, du_3\, (u_2^- u_3^+) (u_2^+ u_3^+)\,
\frac{1}{k^4 R_k (k+p)^4 R_{k+p}}  \qquad\nonumber\\
&& \times\, \frac{(k+p)^4 R_{k+p}- k^4 R_k}{(k+p)^2-k^2}\cdot \frac{(k+p)^2 R_{k+p} - k^2 R_k}{(k+p)^2-k^2}\, \bm{F}_{\mbox{\scriptsize linear}}^{++A}(p,\theta,u_2)\,\bm{V}_{\mbox{\scriptsize linear}}^{--A}(-p,\theta,u_1)\,
\nonumber\\
&& \times \bigg\{  D_1^{--} \delta^{(2,-2)}(u_1,u_2)\, \delta^{(2,-2)}(u_1,u_3) - D_1^{--} \delta^{(2,-2)}(u_1,u_3)\, \delta^{(2,-2)}(u_1,u_2)\bigg\}.
\end{eqnarray}

\noindent
Next, we calculate the integral over $u_1^+$ with the help of the harmonic $\delta$-functions,

\begin{eqnarray}
&&\hspace*{-7mm} -\frac{ie_0^2C_2}{4} \frac{d}{d\ln\Lambda} \int \frac{d^{6}p}{(2\pi)^6} \frac{d^6k}{(2\pi)^6} \int d^8\theta \int du_2\, du_3\,(u_2^- u_3^+) (u_2^+ u_3^+)\, \frac{1}{k^4 R_k (k+p)^4 R_{k+p}}
\nonumber\\
&&\hspace*{-7mm} \times \frac{(k+p)^4 R_{k+p}- k^4 R_k}{(k+p)^2-k^2} \cdot \frac{(k+p)^2 R_{k+p} - k^2 R_k}{(k+p)^2-k^2}\, \bm{F}_{\mbox{\scriptsize linear}}^{++A}(p,\theta,u_2)\,
\bigg\{\bm{V}_{\mbox{\scriptsize linear}}^{--A}(-p,\theta,u_3)\, \nonumber\\
&&\hspace*{-7mm} \times\, D_3^{--} \delta^{(2,-2)}(u_3,u_2) - \bm{V}_{\mbox{\scriptsize linear}}^{--A}(-p,\theta,u_2)\, D_2^{--} \delta^{(2,-2)}(u_2,u_3)\,\bigg\}.
\end{eqnarray}

\noindent
After integrating the harmonic derivatives by parts and keeping in mind that

\begin{equation}\label{U_Identities}
(u^+ u^-) = - (u^- u^+) = 1,\qquad (u^+ u^+) = 0,\qquad (u^- u^-)=0,
\end{equation}

\noindent
this expression can be transformed to the form

\begin{eqnarray}
&&\hspace*{-7mm} -\frac{ie_0^2C_2}{2} \frac{d}{d\ln\Lambda} \int \frac{d^{6}p}{(2\pi)^6} \int d^8\theta\,du\, \bm{V}_{\mbox{\scriptsize linear}}^{--A}(-p,\theta,u)\,
\bm{F}_{\mbox{\scriptsize linear}}^{++A}(p,\theta,u) \int \frac{d^6k}{(2\pi)^6}\, \frac{1}{k^4 R_k (k+p)^4 R_{k+p}}  \nonumber\\
&&\hspace*{-7mm} \times \frac{(k+p)^4 R_{k+p}- k^4 R_k}{(k+p)^2-k^2}\cdot \frac{(k+p)^2 R_{k+p} - k^2 R_k}{(k+p)^2-k^2}.
\end{eqnarray}

\noindent
The integral over the loop momentum $k^\mu$ in the limit $\Lambda\to\infty$ is logarithmically divergent and depends on $p^2/\Lambda^2$.
After differentiating this integral with respect to $\ln\Lambda$ in the limit $\Lambda\to \infty$ (or, equivalently, $p^\mu\to 0$) one obtains a finite constant.
Therefore, the integral over $k^\mu$ should be calculated in the limit of the vanishing external momentum (and obviously after performing Wick rotation).
The corresponding Euclidean momentum will be denoted by $K^\mu$. Thus, the  contribution to the expression (\ref{Expression_For_Calculating}) considered is given by

\begin{equation}
- \frac{e_0^2 C_2}{2}  \int d\zeta^{(-4)} (\bm{F}_{\mbox{\scriptsize linear}}^{++A})^2 \frac{d}{d\ln\Lambda} \int \frac{d^6K}{(2\pi)^6}\, \frac{(K^4 R_K)'(K^2 R_K)'}{K^8 R_K^2},
\end{equation}

\noindent
where primes mark the derivatives with respect to $K^2$.

It is also necessary to add the corresponding contribution coming from the supergraph with a loop of the Pauli--Villars superfield.
By construction, the vertices containing the Pauli--Villars superfield are the same as the ones quadratic in the quantum gauge superfield $v^{++}$,
while the propagator is different due to the presence of the mass term. Moreover, each loop of the Pauli--Villars superfield contributes a minus sign,
see Sect. \ref{Subsection_Quantization}. Therefore, the Pauli--Villars contribution can be obtained through replacing
the momentum part of the propagator as

\begin{equation}\label{PV_Replacement}
\frac{1}{K^4 R_K} \to \frac{1}{K^4 R_K + M^4},
\end{equation}

\noindent
as well as changing the overall sign. Then the considered contribution (corresponding to the first term in eq. (\ref{Vertex_3})) takes the form

\begin{equation}
- C_2\,\mbox{tr} \int d\zeta^{(-4)} (\bm{F}^{++}_{\mbox{\scriptsize linear}})^2 \frac{d}{d\ln\Lambda} \int \frac{d^6K}{(2\pi)^6}  \bigg\{\frac{(K^4 R_K)'(K^2 R_K)'}{K^8 R_K^2}
- \frac{(K^4 R_K)'(K^2 R_K)'}{(K^4 R_K + M^4)^2}\bigg\}.
\end{equation}

The contributions of other terms in the expression (\ref{Vertex_3}) are obtained similarly, but for calculating some of them it is also necessary to use the identity

\begin{equation}
\delta^8(\theta_1-\theta_2) (D_1^+)^4 (D_2^+)^4 (D_3^+)^4 \delta^{14}(z_1-z_2) = (u_1^+ u_2^+)^2 (u_1^+ u_3^+)^2 (u_2^+ u_3^+)^2 \partial^2 \delta^{14}(z_1-z_2).
\end{equation}

\noindent
The sum of all contributions is given by the expression

\begin{eqnarray}\label{2B_Contribution}
&& \Delta_{(2b)} = C_2\, \mbox{tr} \int d\zeta^{(-4)} (\bm{F}^{++}_{\mbox{\scriptsize linear}})^2 \frac{d}{d\ln\Lambda} \int \frac{d^6K}{(2\pi)^6}\,
\bigg\{- \frac{(K^4 R_K)' (K^2 R_K)'}{K^8 R_K^2}  - \frac{(K^4 R_K)'}{K^8 R_K} \qquad \nonumber\\
&& + \frac{(K^4R_K)' (K^2 R_K)'}{(K^4 R_K + M^4)^2} + \frac{(K^4 R_K)' R_K}{(K^4 R_K + M^4)^2}\bigg\}.
\end{eqnarray}

Next, it is necessary to calculate the supergraph (5b) in Fig. \ref{Figure_2Point_Subdiagrams}. The corresponding vertex is given by eq. (\ref{Vertex_4}).
Once again, for simplicity, we will assume that a loop corresponds to the quantum gauge superfield $v^{++}$. In order to avoid the appearance of ill-defined expressions,
we present the vertex in the form

\begin{eqnarray}
&&\hspace*{-5mm} \pm \frac{e_0^2}{4} f^{ABC} f^{CDE}\lim\limits_{u_3\to u_1}\int d^8\theta\, du_1\, du_2\, \frac{(u_1^- u_2^+)}{(u_1^+ u_2^+)^3} \,(D_1^+)^4 v^{++A}(-k,\theta,u_2)
\bm{F}_{\mbox{\scriptsize linear}}^{++B}(0,\theta,u_1) \qquad\nonumber\\
&&\hspace*{-5mm} \times\, \bm{V}_{\mbox{\scriptsize linear}}^{--D}(0,\theta,u_1) \Big((k^2 R_k)'' + R_k'\Big) D_3^{--} v^{++E}(k,\theta,u_3) + \ldots,
\end{eqnarray}

\noindent
where  dots denote the terms which give zero contributions to the tadpole graph. Then the corresponding part of the expression (\ref{Expression_For_Calculating})
obtained with the help of the Feynman rules is written as

\begin{eqnarray}\label{5B_First_Term_Contribution}
&& -\frac{ie_0^2 C_2}{2}\lim\limits_{u_3\to u_1} \int \frac{d^6p}{(2\pi)^6}  \int d^8\theta\, du_1\, du_2\, \bm{V}_{\mbox{\scriptsize linear}}^{--A}(-p,\theta,u_1)
\bm{F}_{\mbox{\scriptsize linear}}^{++A}(p,\theta,u_1) \frac{(u_1^- u_2^+)}{(u_1^+ u_2^+)^3}\, (D_1^+)^4 \qquad\nonumber\\
&& \times (D_2^+)^4 \delta^8(\theta_1-\theta_2)\Big|_{\theta_1=\theta_2=\theta} D_3^{--} \delta^{(-2,2)}(u_2,u_3) \frac{d}{d\ln\Lambda} \int \frac{d^6k}{(2\pi)^6} \frac{(k^2 R_k)''+R_k'}{k^4 R_k}.
\end{eqnarray}

\noindent
Note that here we took into account that, when calculating the derivative of the integral over the loop momentum $k^\mu$ with respect to $\ln\Lambda$ in the limit $\Lambda\to \infty$,
the external momentum $p^\mu$ in the integral under consideration {\it should} be set equal to 0.

Taking into account that

\begin{equation}
(D_1^+)^4 (D_2^+)^4 \delta^8(\theta_1-\theta_2)\Big|_{\theta_1=\theta_2=\theta} = (u_1^+ u_2^+)^4,
\end{equation}

\noindent
the expression (\ref{2B_First_Term_Contribution}) can be rewritten  as

\begin{eqnarray}
&&\hspace{-5mm} -\frac{ie_0^2C_2}{2} \int \frac{d^6p}{(2\pi)^6}  \int d^8\theta\, du_1\, du_2\, \bm{V}_{\mbox{\scriptsize linear}}^{--A}(-p,\theta,u_1) \bm{F}_{\mbox{\scriptsize linear}}^{++A}(p,\theta,u_1)\nonumber\\
&&\hspace{-5mm}\qquad\qquad\qquad \times (u_1^- u_2^+)(u_1^+ u_2^+)\, D_1^{--} \delta^{(-2,2)}(u_2,u_1) \frac{d}{d\ln\Lambda} \int \frac{d^6k}{(2\pi)^6} \frac{(k^2 R_k)''+R_k'}{k^4 R_k}.\qquad
\end{eqnarray}

\noindent
Integrating by parts with respect to the derivative $D_1^{--}$, making use of the identities (\ref{U_Identities}), and calculating the harmonic integral over $u_2$
with the help of the harmonic $\delta$-function, we cast the result in the form

\begin{equation}
\frac{ie_0^2C_2}{2} \int \frac{d^6p}{(2\pi)^6}  \int d^8\theta\, du\, \bm{V}_{\mbox{\scriptsize linear}}^{--A}(-p,\theta,u)
\bm{F}_{\mbox{\scriptsize linear}}^{++A}(p,\theta,u) \frac{d}{d\ln\Lambda} \int \frac{d^6k}{(2\pi)^6} \frac{(k^2 R_k)''+R_k'}{k^4 R_k}.
\end{equation}

\noindent
After Wick rotation and adding the Pauli--Villars contribution (which is obtained by the replacement (\ref{PV_Replacement})
and changing the overall sign) we obtain the contribution of the supergraph (5b) in Fig. \ref{Figure_2Point_Subdiagrams} to the expression (\ref{Expression_For_Calculating})

\begin{equation}\label{5B_Contribution}
\Delta_{(5b)} = C_2\,\mbox{tr} \int d\zeta^{(-4)}\, (\bm{F}_{\mbox{\scriptsize linear}}^{++})^2 \frac{d}{d\ln\Lambda} \int \frac{d^6K}{(2\pi)^6} \bigg\{\frac{(K^2 R_K)''+ R_K'}{K^4 R_K}
- \frac{(K^2 R_K)''+R_K'}{K^4 R_K + M^4}\bigg\}.
\end{equation}

Summing up the expressions (\ref{2B_Contribution}) and (\ref{5B_Contribution}), we obtain the contribution to the expression (\ref{Expression_For_Calculating})
coming from the one-loop superdiagrams (2b) and (5b) in Fig. \ref{Figure_2Point_Subdiagrams},

\begin{eqnarray}
&&\hspace*{-3mm} \frac{d}{d\ln\Lambda} \Delta\Gamma^{(2)}_{\bm{V}^{++},\infty} = C_2\, \mbox{tr} \int d\zeta^{(-4)} (\bm{F}^{++}_{\mbox{\scriptsize linear}})^2\,
\frac{d}{d\ln\Lambda} \int \frac{d^6K}{(2\pi)^6} \bigg\{-\frac{4}{K^6} - \frac{R_K'}{K^4 R_K} - \frac{(R_K')^2}{K^2 R_K^2}  \nonumber\\
&&\hspace*{-3mm} + \frac{R_K''}{K^2 R_K} + \frac{4 K^2 R_K^2 + 4 K^4 R_K R_K' + K^6 (R_K')^2}{(K^4 R_K + M^4)^2} - \frac{K^2 R_K'' + 3R_k'}{K^4 R_K + M^4}\bigg\} + O(\alpha_0).
\end{eqnarray}

\noindent
It is convenient to split this expression into two parts. One of them does not contain the mass $M$ and comes from the supergraphs with a loop
of the quantum gauge superfield $v^{++}$. The other part contains terms with massive propagators and comes from the superdiagrams involving a loop of the
Pauli--Villars superfield $P^{++}$. Like in the case of $4D$, ${\cal N}=1$ supersymmetric theories regularized by higher covariant derivatives \cite{Aleshin:2016yvj},
it is possible to show that each of these two parts can be written as an integral of the double total derivative,

\begin{eqnarray}
&& \frac{d}{d\ln\Lambda} \Delta\Gamma^{(2)}_{\bm{V}^{++},\infty} = \frac{C_2}{4}\, \mbox{tr} \int d\zeta^{(-4)} (\bm{F}^{++}_{\mbox{\scriptsize linear}})^2\,
\frac{d}{d\ln\Lambda} \int \frac{d^6K}{(2\pi)^6}\,\frac{\partial^2}{\partial K_\mu\,\partial K^\mu}
\nonumber\\
&&\qquad\qquad\qquad\qquad\qquad \times \bigg\{\frac{1}{K^4} \ln (K^4 R_K) - \frac{1}{K^4}\ln(K^4 R_K + M^4)\bigg\} + O(\alpha_0).\qquad
\end{eqnarray}

\end{document}